\documentclass[10pt,twocolumn,letterpaper]{article}
\usepackage[pagenumbers]{cvpr}
\usepackage[utf8]{inputenc}
\usepackage[T1]{fontenc}
\usepackage{microtype}
\usepackage{nicefrac}
\usepackage{multirow}
\usepackage{makecell}
\usepackage{enumitem}
\usepackage{array}
\usepackage{adjustbox}
\newcommand{\fitwidth}[1]{\begin{adjustbox}{max width=\linewidth}#1\end{adjustbox}}
\definecolor{cvprblue}{rgb}{0.21,0.49,0.74}
\usepackage[breaklinks,hidelinks]{hyperref}

\providecommand{\Description}[1]{}

\title{FeatMark: Feature-level Watermark Protection against Mimicry Attacks with Diffusion Models}
\author{Haoyang Li$^{1}$ \quad Ruoxi Sun$^{2}$ \quad Qingqing Ye$^{1}$ \quad Benjamin Zi Hao Zhao$^{3}$\\
Yaxin Xiao$^{1}$ \quad Jason Xue$^{2}$ \quad Haibo Hu$^{1,*}$\\[3pt]
$^{1}$The Hong Kong Polytechnic University \quad $^{2}$CSIRO's Data61\\
$^{3}$Macquarie University\\[3pt]
{\small $^{*}$Corresponding author: \href{mailto:haibo.hu@polyu.edu.hk}{haibo.hu@polyu.edu.hk}}
}
\hypersetup{pdftitle={FeatMark: Feature-level Watermark Protection against Mimicry Attacks with Diffusion Models},pdfauthor={Haoyang Li, Ruoxi Sun, Qingqing Ye, Benjamin Zi Hao Zhao, Yaxin Xiao, Jason Xue, Haibo Hu}}

\begin{document}
\raggedbottom

\maketitle

\begin{abstract}
Text-to-image diffusion models enable data-efficient ``mimicry'' attacks, wherein adversaries fine-tune the model on a handful of public photos to synthesize convincing forgeries of a target individual. A common countermeasure is to embed imperceptible, low-energy watermarks, yet recent studies show these signatures are brittle: modest post-processing or lightweight adversarial perturbations readily suppress detection, exposing a fundamental tension between imperceptibility and robustness. We introduce \textbf{FeatMark}, a watermarking framework that shifts from pixel-level, energy-starved perturbations to \emph{inconspicuous semantic features}: small, scene-consistent micro-features that remain natural to humans while providing a stronger, machine-verifiable provenance signal. FeatMark builds domain-specific feature banks that encode each watermark as a compact concept program, pairing open-vocabulary semantic cues with reliable edit regions and instruction templates. It then automatically selects features that are both feasible and executable and injects them through modular, mask-guided concept editing, yielding highly localized, scene-consistent micro-edits that are difficult to perceive.
We conduct extensive experiments across VGGFace2, CelebA-HQ, and WikiArt, evaluating against 10 strong watermark \emph{removal/purification} attacks (including regeneration-style purification) and several bespoke \emph{adaptive} attacks tailored to FeatMark, to assess perceptual fidelity, watermark detection accuracy, and robustness.
We further demonstrate FeatMark's extensibility to video mimicry attacks.
The results show FeatMark remains virtually impervious, withstanding all evaluated attacks with negligible bit-accuracy and fidelity degradation.

\end{abstract}


\section{Introduction}
\begin{figure}[t]
\centering
\includegraphics[width=\linewidth]{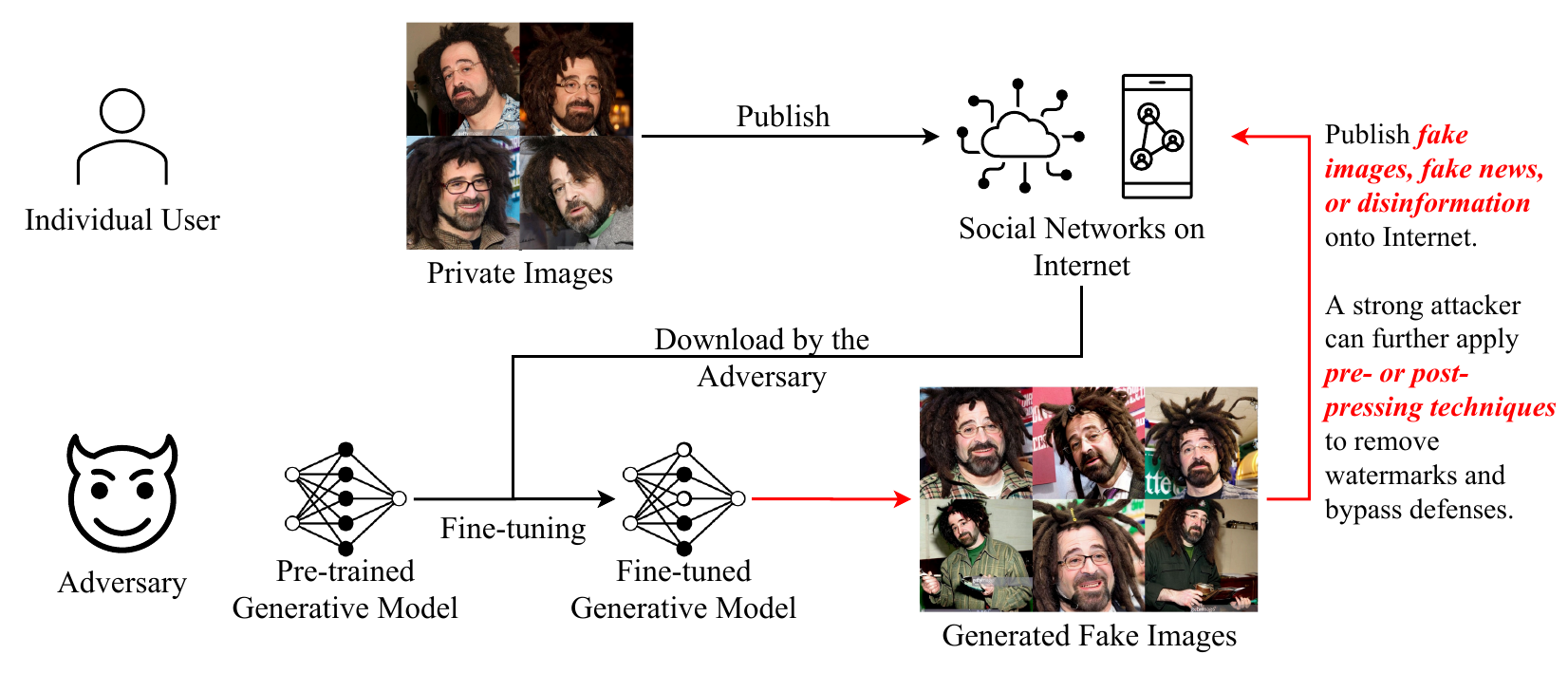}
\caption{An illustration of mimicry attacks.}
\label{fig:mimicry_attack}

\Description{} \end{figure}

Text-to-image diffusion models such as Stable Diffusion~\cite{stable_diffusion} and DALL-E~\cite{dall-e} empower creative synthesis but also make mimicry attacks feasible~\cite{deepfake_review_22, stylegan, diffusion_beats_gan}, in which an adversary fine-tunes a pre-trained generator on just a handful of a victim’s public photos so the model learns to replicate that person’s distinctive appearance~\cite{dreambooth,fine_tuning_diffusion_models_22NIPSw,fine_tuning_diffusion_models_23CVPR}, as Figure~\ref{fig:mimicry_attack} shows.
Watermarks offer a direct countermeasure, embedding persistent provenance signals into those fake images so that every derivative remains traceable.
Recent regulations (\eg EU AI Act Art. 50~\cite{EUAIAct_Art50_2024} and US EO 14110~\cite{EO14110_2023}) encourage reliable watermarking, underscoring the need for robust provenance.
Regarding these regulations, contemporary methods such as Tree-Ring Watermarks~\cite{tree-ring-watermarks} and FT-Shield~\cite{ft-shield} hide imperceptible signatures in attack results by regulating diffusion models or perturbing public images.

However, extant watermarking schemes are demonstrably brittle under adversarial manipulation.
Most state-of-the-art watermarking methods~\cite{hidden, stablesignature, tree-ring-watermarks, diffusionshield, ft-shield} insert low-amplitude noise that is invisible to humans yet statistically detectable by a verifier. Although this preserves visual fidelity, it grants attackers a powerful lever: small, human-invisible adversarial perturbations can overwhelm or erase the signature without perceptual cost~\cite{stable_signature_is_unstable, wevade, transfer_attack_image_watermarks}.
Even worse, the adversarial perturbations required for such evasion are computationally and perceptually cheap. For instance, WEvade~\cite{wevade} achieves perfect bypass with imperceptible $\ell_\infty$-bounded noise in only a few PGD iterations.
These empirical findings reveal a fundamental trade-off: as watermark energy is reduced to maintain imperceptibility, every incremental gain in stealth tends to reduce the watermark’s energy budget for robustness, perpetuating an arm race that defenders cannot win with the current paradigm~\cite{audiomarkbench, guo2024aigeneratedimagedetectionpassive, watermark_impossibility}. Consequently,
it becomes imperative to develop a fundamentally different watermarking mechanism that does not rely solely on imperceptibility and thus cannot be trivially nullified by lightweight adversarial perturbations.

To overcome this impasse, we introduce the concept of {\it inconspicuous-feature} watermarks. Instead of hiding energy-starved pixel perturbations, we affix semantically coherent micro-features (\eg a faint botanical emblem subtly woven into existing foliage) that are perceptible on close inspection yet blend naturally with the generated scene. Their greater signal strength lets them withstand adversarial perturbations while still escaping casual human notice.
This principle has clear precedents in security. Patch-based adversarial examples use naturalistic artifacts to mislead detectors~\cite{ae_survey, natural_ae, chen2023advdiffuser}, and semantic backdoors leverage real objects to control model outputs~\cite{bagdasaryan2020backdoor, sun2024neural, wang2023versatile}. These generative-prior features often go unnoticed yet remain effective, motivating our inconspicuous-feature strategy.
More importantly, in the scenario of mimicry attack, stealth of inconspicuous feature is aided by an identity-based knowledge gap and diffusion stochasticity~\cite{hallucination_llm}, which make the features hard to anticipate and easy to ascribe to sampling variance.

Building on these insights, we propose FeatMark, a fully automatic framework that injects inconspicuous semantic micro-features into images and transfers them through mimicry training via image cloaking.
FeatMark first constructs domain-specific \emph{feature banks}, where each entry is a compact \emph{concept program} that bundles open-vocabulary semantic cues (for recognizing concept evidence), a reliable region-of-edit specification, and instruction templates that a generic editor can follow.
Given a target identity or artistic style, FeatMark automatically scores candidate concepts using open-vocabulary recognition and selects a small set of executable features, then performs mask-guided concept-based local editing to realize localized and scene-consistent micro-edits.
Crucially, FeatMark does not publish the edited images directly. Instead, it produces the released images by cloaking the original images toward their edited counterparts, so that the published set remains visually faithful while behaving like the watermarked set under mimicry training representations.
Finally, a private watermark reader is trained to decode a user-specific message from suspicious images, enabling attribution of mimicry outputs without requiring access to the attacker’s model.

We evaluate FeatMark across both facial and artistic domains, including VGGFace2~\cite{vggface2}, CelebA-HQ~\cite{celebahq1}, and WikiArt~\cite{wikiart}.
Across datasets, FeatMark preserves perceptual quality close to the no-defense baseline while providing a verifiable watermark signal with high decoding accuracy on watermarked images and near-random decoding on clean images.
We further test robustness against a broad suite of strong purification and watermark removal attacks, including IMPRESS~\cite{impress}, Noisy Upscaling~\cite{adversarial_perturbations_cannot}, WEvade~\cite{wevade}, UnMarker~\cite{unmarker}, regeneration-style removals (diffusion/denoise~\cite{diffpure} and VAE~\cite{vae}), and WatermarkAttacker~\cite{watermark_attacker}.
In settings where pixel-level invisible watermarks collapse, FeatMark’s decoding remains essentially unchanged under most attacks (typically a $\le 0.02$ drop), and even under stronger regeneration-style purification the accuracy degradation largely coincides with visible fidelity loss, leaving FeatMark still decodable.
To probe future, attacker-tailored strategies, we additionally design bespoke adaptive attacks targeted to FeatMark’s pipeline and show that removing the watermark without noticeable identity/style drift remains difficult.
Finally, we demonstrate that the same feature-anchored design transfers to video mimicry.

To summarize, our key contributions are as follows:

\begin{itemize}[leftmargin=*]
\item To our knowledge, FeatMark is the \textit{first} watermarking scheme that protects against mimicry attacks by anchoring provenance in inconspicuous semantic micro-features rather than energy-starved pixel perturbations.
\item We introduce a fully automatic watermarking pipeline built on feature-bank concept programs, automatic executable feature selection, and mask-guided concept-based local editing, coupled with a cloaking stage that enables safe publishing while preserving watermark transfer through mimicry training.
\item We provide comprehensive evaluations over perceptual fidelity, watermark decoding accuracy, and robustness to strong purification, removal, and regeneration attacks, and further assess attacker-tailored adaptive attacks designed for FeatMark.
\end{itemize}

\section{Background}
\label{sec:background}

\noindent\textbf{Mimicry attacks.}
We study \emph{mimicry attacks} where an adversary exploits a small set of publicly released personal images of a target individual, denoted as $\mathcal{X}=\{x\}$, to fine-tune a generative model and synthesize high-fidelity impersonating images. Following prior work on unauthorized personalization and style/identity mimicry~\cite{impress,glaze,dreambooth}, we model the attacker as fine-tuning a text-to-image diffusion model $\mathcal{M}^{\theta}$ and sampling fake images $x_{\text{fake}}$ conditioned on prompt $c$. The attacker optimizes:
\begin{equation}
\theta^* = \arg\max_{\theta}\ \mathbb{E}_{\substack{x_{\text{fake}}\sim\,p(x_{\text{fake}}\mid c,\mathcal{M}^{\theta})}}\!\big[\mathcal{A}(x_{\text{fake}},\mathcal{X})\big],
\label{eq:attack_obj}
\end{equation}
where $\mathcal{A}$ is an attacker-chosen similarity/utility metric that encourages the synthesized outputs to match the target identity or style.

\noindent\textbf{Attribution via watermarking.}
Our goal is \emph{attribution} rather than disruption: the defender transforms their released images such that mimicry outputs tend to carry a user-specific signature that can later be verified. Concretely, the defender holds a private message $m$ and a watermark reader $\mathcal{R}$ that decodes a bit sequence from a suspicious image. We aim for high expected bit agreement between $m$ and $\mathcal{R}(\cdot)$ on attacker-generated samples:
\begin{equation}
\begin{aligned}
\max\ & \mathbb{E}_{x_{\text{fake}}\sim p(x_{\text{fake}}\mid c,\mathcal{M}^{\theta^*})}
\big[\mathcal{B}_{\text{acc}}(m,\mathcal{R}(x_{\text{fake}}))\big] \\
\text{s.t. }\ & \theta^* \text{ satisfies Eq.~\ref{eq:attack_obj}}.
\end{aligned}
\label{eq:watermarking}
\end{equation}
Here, $\mathcal{B}_{\text{acc}}$ measures the fraction of correctly recovered bits.

\noindent\textbf{Adaptive attackers.}
A defender-facing evaluation must consider adaptive attackers who post-process outputs to evade verification while preserving identity/style. We model such post-processing by an attacker-side transformation $\mathcal{P}$, yielding the adaptive objective:
\begin{equation}
\begin{aligned}
\theta^* = \arg\max_{\theta}\;& \mathbb{E}_{\substack{x_{\text{fake}}\sim\,p(x_{\text{fake}}\mid c,\mathcal{M}^{\theta})}}\!\big[\mathcal{A}(\mathcal{P}(x_{\text{fake}}),\mathcal{X})\big].
\end{aligned}
\label{eq:adaptive_obj}
\end{equation}
This motivates designing watermarking mechanisms that are robust to unknown, potentially strong adaptive strategies.

\begin{figure*}[t]
\centering
\includegraphics[width=1.0\linewidth]{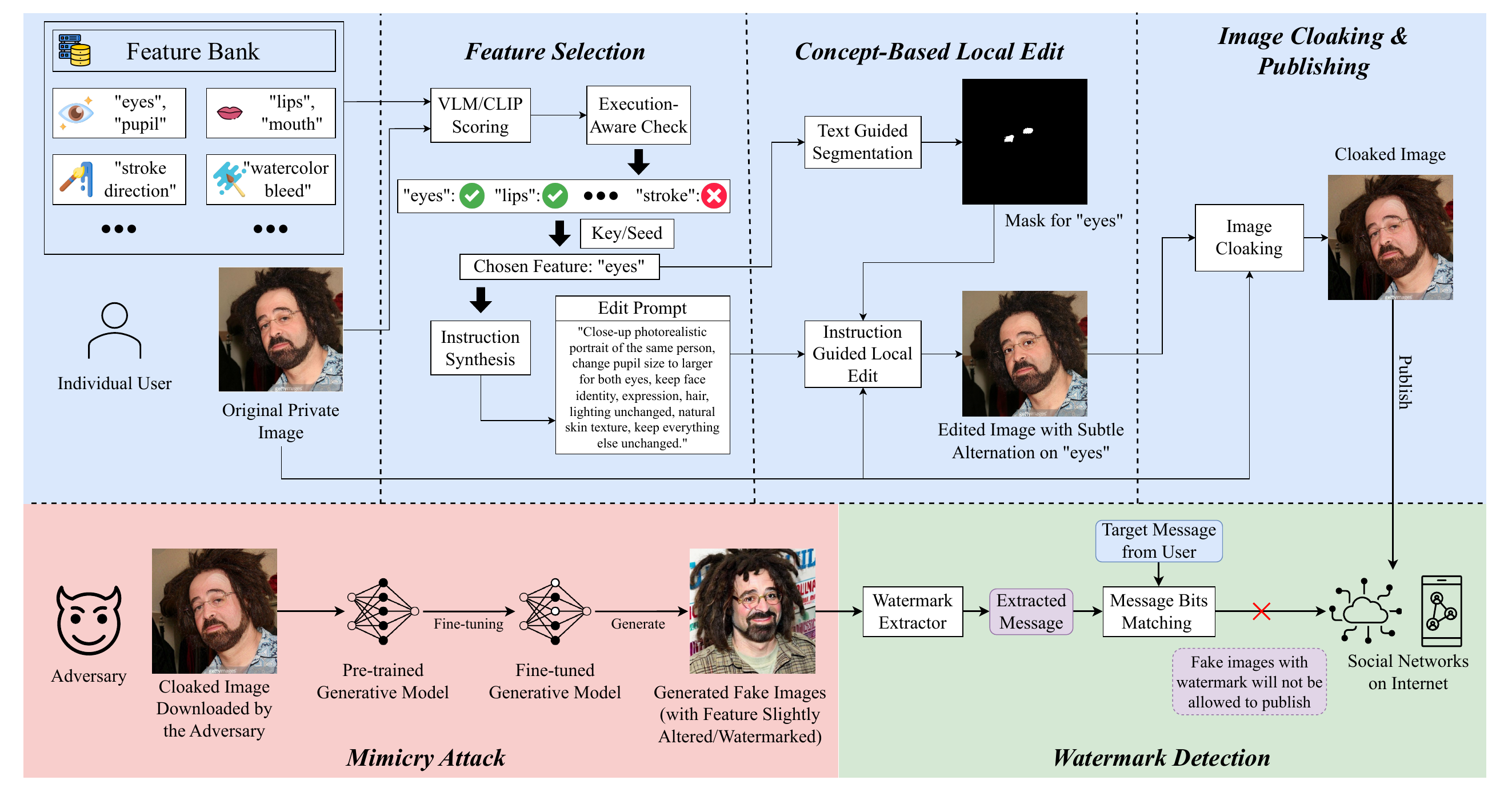}
\caption{FeatMark overview. Given a small set of personal images, FeatMark queries a curated Feature Bank of concept programs and automatically selects an executable feature using open-vocabulary VLM/CLIP scoring and execution-aware checks. It then localizes an editable region via text-guided grounding and performs instruction-guided local editing to produce subtly watermarked images. Perturbing the original images with the edited images using image cloak, the cloaked images will then be published to Internet. The attacker can only perform mimicry attacks on the published images and generate fake images. Finally, the watermark extractor will compare the extracted message with the target message that is provided by the user to identify watermarked fake images and stop them being published online.}
\label{fig:overview}

\Description{}
\end{figure*}

\section{Methodology}
\label{sec:methodology}

This section specifies the threat model, summarizes the end-to-end workflow, and details each module of FeatMark.

\subsection{Threat Model}
\label{sec:threat_model_problem_statement}

We follow threat models in prior work on protecting personal data from malicious personalization~\cite{glaze,photoguard,anti-dreambooth,advdm}. The defender is an individual user who publishes a small set of images and later wishes to attribute whether a suspicious image was generated by a mimicry model trained on those released images. The defender holds a private target message $m$ and runs a private watermark reader $\mathcal{R}$ locally, which are not exposed to the attacker.

The attacker has access to the published images and sufficient compute to fine-tune a text-to-image diffusion model $\mathcal{M}^{\theta}$ to generate impersonating outputs, as in Eq.~\ref{eq:attack_obj}. We assume the attacker knows the defense algorithm (including the existence of feature-bank-driven watermarking), and may deploy advanced strategies such as purification $\mathcal{P}$ or regeneration-based transformations to evade verification, but does not have access to the defender’s private verification parameters.

\subsection{Overview of Our Watermark Workflow}
\label{sec:overview}

FeatMark is a fully automatic pipeline that converts a small personal image set into publishable training data that transfers a user-specific semantic watermark through mimicry training. Starting from the original set $\mathcal{X}=\{x\}$, FeatMark queries a Feature Bank $\mathcal{F}=\{f_k\}_{k=1}^{K}$ whose entries are executable \emph{concept programs}. Each concept program bundles textual descriptors for open-vocabulary concept evidence, grounding prompts for localizing an editable region, and instruction templates for expressing the concept via a generic editor. This design removes the need for manual feature crafting and manual masking.

Given $\mathcal{X}$, FeatMark automatically selects a feature that is semantically suitable for the set and operationally executable. It then performs localized, instruction-guided concept editing to obtain a privately held watermarked set $\mathcal{X}^{\mathrm{wm}}=\{x^{\mathrm{wm}}\}$, where each $x^{\mathrm{wm}}$ differs subtly from its corresponding $x$ while reliably expressing the selected concept.

Crucially, the images that are actually published are produced by an image cloaking stage. For each original image $x$, FeatMark computes a released image $x^{\mathrm{pub}}$ by adding an imperceptible perturbation so that $x^{\mathrm{pub}}$ remains visually close to $x$ but is optimized to behave like $x^{\mathrm{wm}}$ under the representations used by mimicry training, yielding a released set $\mathcal{X}^{\mathrm{pub}}=\{x^{\mathrm{pub}}\}$. Only $\mathcal{X}^{\mathrm{pub}}$ is shared with the public and thus accessible to an attacker for fine-tuning, while $\mathcal{X}^{\mathrm{wm}}$ is kept private and later used as supervision. This cloaking step follows the perturbation-based protection paradigm in prior work~\cite{glaze,photoguard} but is integrated here to facilitate semantic watermark transfer rather than merely disrupting generation.

After generating the publishable set, the defender trains a private watermark reader $\mathcal{R}$ using a small number of clean images from $\mathcal{X}$ and their corresponding privately stored watermarked counterparts from $\mathcal{X}^{\mathrm{wm}}$. At test time, given a suspicious image, the defender decodes a bit sequence with $\mathcal{R}$ and verifies agreement with the user’s private message $m$, enabling attribution of mimicry outputs generated from models fine-tuned on $\mathcal{X}^{\mathrm{pub}}$.

\subsection{Feature Bank Construction}
\label{sec:feature-bank}

FeatMark builds on a feature bank, a library of watermarkable concepts that can be consistently recognized from an image and later reproduced by a mimic generator.
Each entry $f_k \in \mathcal{B}$ is specified as a compact concept program that defines how the concept is queried and described.
Concretely, $f_k$ provides a small set of textual queries $\mathcal{T}_k=\{t_{k,1},\dots,t_{k,J}\}$ (paraphrases/templates) that all refer to the same concept, and a canonical name $\text{name}(f_k)$ that will later be used to form an editing instruction.
The guiding intuition is to treat a feature as an atomic semantic attribute whose presence leaves a stable trace in representation space, while remaining editable with minimal perceptual disruption.

To measure concept evidence, we use an open-vocabulary vision-language model~\cite{clip} to map an image $x$ and a text query $t$ into a joint embedding space.
We define the evidence score of feature $f_k$ on image $x$ as
\begin{equation}
s_k(x) \;=\; \max_{t \in \mathcal{T}_k} \ \cos\!\big(\phi_I(x),\, \phi_T(t)\big),
\label{eq:concept-score}
\end{equation}
where $\phi_I(\cdot)$ and $\phi_T(\cdot)$ denote the image and text encoders, and $\cos(\cdot,\cdot)$ is cosine similarity.
Taking the maximum over paraphrases follows the standard prompt-ensembling practice for reducing prompt sensitivity in open-vocabulary recognition \cite{clip,zhou2022cocoop}.
Importantly, Eq.~\eqref{eq:concept-score} only requires a generic VLM interface and does not rely on a specific backbone.

Our bank construction follows two principles with theoretical support from feature selection and representation learning.
First, each feature's evidence should be stable across natural variations of the same identity or style, which encourages downstream generators to internalize and reuse the cue.
This is aligned with the well-documented tendency of modern models to exploit stable, high-signal shortcuts when such cues remain predictive in-distribution \cite{geirhos2020shortcut}.
Second, the bank should be diverse and non-redundant, since near-duplicate concepts waste watermark capacity and make automatic selection brittle.
We therefore curate $\mathcal{B}$ to cover multiple semantic axes and later enforce redundancy control during selection using classical relevance, redundancy criteria from information-theoretic feature selection \cite{guyon2003introduction,peng2005mrmr,brown2012itfs}.
These principles keep the bank compact, interpretable, and amenable to fully automatic downstream selection.

\subsection{Automatic Feature Selection from the Bank}
\label{sec:feature-selection}

Given a target set of images $\mathcal{X}=\{x_i\}_{i=1}^n$ for one identity or style, FeatMark selects a small subset of features $\mathcal{K}\subset \{1,\dots,|\mathcal{B}|\}$ without any manual screening.
The selection aims to identify concepts that are consistently detectable on $\mathcal{X}$ and also complementary to each other, so that watermark bits are distributed over multiple independent semantic traces rather than concentrated on a single fragile cue.

We first compute evidence statistics for each bank feature using Eq.~\eqref{eq:concept-score}.
For feature $f_k$, we aggregate scores across the set,
\begin{equation}
\begin{aligned}
\mu_k &= \frac{1}{n}\sum_{i=1}^n s_k(x_i), \\
\sigma_k^2 &= \frac{1}{n}\sum_{i=1}^n \big(s_k(x_i)-\mu_k\big)^2,
\end{aligned}
\label{eq:score-stats}
\end{equation}
and keep candidates with sufficiently high mean evidence and low variance.
This filter-style step favors stable cues over spurious ones and follows standard practice in feature selection \cite{guyon2003introduction}.
To avoid selecting multiple semantically overlapping features, we further apply redundancy-aware pruning.
Let $\psi_k$ denote a textual embedding of $\text{name}(f_k)$ (or the centroid of $\{\phi_T(t): t\in\mathcal{T}_k\}$). 
We penalize candidates whose textual representations are too similar to already selected ones and greedily grow $\mathcal{K}$ by maximizing a relevance--redundancy trade-off, an unsupervised analogue of mRMR-style criteria \cite{peng2005mrmr,brown2012itfs}.

Because this procedure depends only on the bank and $\mathcal{X}$, it enables a fully automatic pipeline while still producing a small set of semantically distinct, high-confidence concepts.
In practice, the resulting $\mathcal{K}$ reduces downstream editing failures and improves robustness by preventing the watermark from collapsing into a single easily targeted attribute.

\subsection{Concept-Based Image Editing}
\label{sec:concept-editing}

Once a feature $f_k$ is selected, FeatMark performs concept-based editing to inject or modify the corresponding semantic cue while keeping the output realistic and identity/style-consistent.
The concept program provides a grounding phrase $\text{ground}(f_k)$ that specifies where the concept should be edited (e.g., an attribute-associated region), and an instruction template $\text{inst}(f_k,\cdot)$ that specifies how it should be edited (e.g., add or strengthen the concept).

We obtain a region-of-edit mask by applying an open-world grounding module to localize $\text{ground}(f_k)$ on image $x$, producing a soft mask $M_k(x)\in[0,1]^{H\times W}$.
The instruction is instantiated as $u_k=\text{inst}(f_k,\text{name}(f_k))$.
A generic concept-guided editor $\mathcal{E}$ then produces an edited image
\begin{equation}
x_k^{\text{edit}} \;=\; \mathcal{E}\!\big(x,\, u_k,\, M_k(x)\big).
\label{eq:edit}
\end{equation}
Our methodology only requires $\mathcal{E}$ to support localized, instruction-following edits, and it can be instantiated by a wide range of modern image editing frameworks, including diffusion-based editors \cite{meng2022sdedit,hertz2022prompttoprompt,couairon2022diffedit,brooks2023instructpix2pix}.

To discourage unnecessary changes outside the edited region, we use a mask-weighted consistency objective during editing (or as a post-selection validation),
\begin{equation}
\begin{aligned}
\mathcal{L}_{\text{out}}(x, x_k^{\text{edit}})
&= \big\|(1-M_k(x)) \odot (x_k^{\text{edit}}-x)\big\|_1.
\end{aligned}
\label{eq:outside-loss}
\end{equation}
This formulation makes the watermark signal semantic: bits correspond to the presence or strength of concepts that are edited in a localized and visually plausible manner, which later yields robustness against perturbation-level removals.

\subsection{Image Cloak with Refined Objective}
\label{method:image_cloak_refined}

In FeatMark, image cloaking is a required stage for producing publishable images. After concept-based editing yields a watermarked set $\mathcal{X}^{\mathrm{wm}}=\{x^{\mathrm{wm}}\}$, we further transform each original image $x$ into a \emph{released} image $x^{\mathrm{pub}}$ that remains visually faithful to $x$ while being optimized to transfer the watermark feature through mimicry training. Concretely, only $\mathcal{X}^{\mathrm{pub}}=\{x^{\mathrm{pub}}\}$ is published and thus accessible to the attacker for fine-tuning, whereas $\mathcal{X}^{\mathrm{wm}}$ is retained privately and will be used in the subsequent watermark reader training as supervision.

Following perturbation-based cloaking methods that add imperceptible noise to influence downstream generative behavior~\cite{glaze,photoguard}, we parameterize the released image as $x^{\mathrm{pub}} = x + \delta$ and optimize $\delta$ so that $x^{\mathrm{pub}}$ stays close to $x$ in pixel space yet aligns with the edited target $x^{\mathrm{wm}}$ in a representation space. Specifically, we refine the standard cloaking objective with a reconstruction-consistency regularizer to reduce overfitting to a single encoder and to avoid degenerate perturbations:
\begin{equation}
\begin{aligned}
\mathcal{L}_{\text{cloak}} ={}& \left\|\mathfrak{E}(x+\delta)- \mathfrak{E}(x^{\text{wm}})\right\|_2^2 \\
&+ \lambda \left\|(x+\delta)- \mathfrak{D}(\mathfrak{E}(x+\delta))\right\|_2^2, \\
\text{s.t.}\quad &\|\delta\|_1 \leq \eta,\quad \forall \delta \in \{\delta\},
\end{aligned}
\label{eq:loss_cloak}
\end{equation}
where $\mathfrak{E}(\cdot)$ and $\mathfrak{D}(\cdot)$ denote the encoder and decoder, respectively. The first term encourages the released image to be close to the watermarked target in the encoder space, which promotes watermark feature transfer when the attacker fine-tunes on $\mathcal{X}^{\mathrm{pub}}$. The second term regularizes the perturbation to remain stable under an encode--decode cycle, mitigating brittle solutions that can arise from purely feature-space alignment~\cite{adversarial_perturbations_cannot,impress}.

\begin{table}[t]
\small
\setlength{\tabcolsep}{2.5pt}
\centering
\caption{Results on perceptual quality for all datasets. For each metric, we calculate the results between fake images affected by the corresponding method and original images. Prec., Rec., Cov., and Dens. denote precision, recall, coverage, and density.}
\label{table:results_perceptual_all}
\fitwidth{
\begin{tabular}{lccccc}
\toprule
\textbf{Method} & \textbf{FID}$\downarrow$ & \textbf{Prec.}$\uparrow$ & \textbf{Rec.}$\uparrow$ & \textbf{Cov.}$\uparrow$ & \textbf{Dens.}$\uparrow$ \\
\midrule
\multicolumn{6}{c}{\textbf{VGGFace2}} \\
No-Defense & 171.81 & 0.65 & 0.28 & 1.33 & 0.83 \\
\cmidrule(r){1-6}
FeatMark & 180.98 & 0.47 & 0.19 & 0.83 & 0.62 \\
\cmidrule(r){1-6}
MetaCloak & 457.10 & 0.00 & 0.00 & 0.00 & 0.00 \\
\cmidrule(r){1-6}
Glaze & 325.89 & 0.01 & 0.00 & 0.01 & 0.01 \\
\cmidrule(r){1-6}
Anti-DreamBooth & 449.74 & 0.00 & 0.00 & 0.00 & 0.00 \\

\midrule
\multicolumn{6}{c}{\textbf{CelebA-HQ}} \\
No-Defense & 128.99 & 0.54 & 0.27 & 1.00 & 0.68 \\
\cmidrule(r){1-6}
FeatMark & 142.33 & 0.31 & 0.18 & 0.44 & 0.41 \\
\cmidrule(r){1-6}
MetaCloak & 437.92 & 0.00 & 0.00 & 0.00 & 0.00 \\
\cmidrule(r){1-6}
Glaze & 279.13 & 0.01 & 0.00 & 0.03 & 0.01 \\
\cmidrule(r){1-6}
Anti-DreamBooth & 465.37 & 0.00 & 0.00 & 0.00 & 0.00 \\

\midrule
\multicolumn{6}{c}{\textbf{WikiArt}} \\
No-Defense & 306.02 & 0.69 & 0.13 & 0.68 & 0.46 \\
\cmidrule(r){1-6}
FeatMark & 321.76 & 0.62 & 0.07 & 0.36 & 0.29 \\
\cmidrule(r){1-6}
MetaCloak & 458.61 & 0.00 & 0.00 & 0.00 & 0.00 \\
\cmidrule(r){1-6}
Glaze & 410.93 & 0.00 & 0.00 & 0.00 & 0.00 \\
\cmidrule(r){1-6}
Anti-DreamBooth & 447.03 & 0.00 & 0.00 & 0.00 & 0.00 \\

\bottomrule
\end{tabular}
}
\end{table}

\begin{table}[t]
\small
\setlength{\tabcolsep}{2.5pt}
\centering
\caption{Bit accuracy for watermark methods on all datasets with/without watermarks and applied with multiple transformations. The column of {\it WM?} denotes whether the evaluated images contain the watermark or not.}
\label{table:results_bit_acc_all}
\fitwidth{
\begin{tabular}{lccccccc}
\toprule
\multirow{2}{*}{\textbf{Method}} 
& \multirow{2}{*}{\textbf{WM?}} 
& \multicolumn{6}{c}{\textbf{Bit Accuracy}} \\
\cmidrule(r){3-8}
& & None & Crop & Brigh. & Cont. & JPEG & Comb. \\
\midrule
\multicolumn{8}{c}{\textbf{VGGFace2}} \\
\multirow{2}{*}{\shortstack{Stable\\Signature}} & Y & 0.99 & 0.36 & 0.98 & 0.97 & 0.66 & 0.80 \\
& N & 0.45 & 0.40 & 0.49 & 0.49 & 0.46 & 0.46 \\
\cmidrule(r){1-8}
\multirow{2}{*}{HiDDeN} & Y & 0.99 & 0.58 & 0.97 & 0.94 & 0.83 & 0.62 \\
& N & 0.43 & 0.47 & 0.42 & 0.45 & 0.41 & 0.45 \\
\cmidrule(r){1-8}
\multirow{2}{*}{StegaStamp} & Y & 0.98 & 0.54 & 0.98 & 0.97 & 0.92 & 0.85 \\
& N & 0.45 & 0.46 & 0.42 & 0.42 & 0.42 & 0.47 \\
\cmidrule(r){1-8}
\multirow{2}{*}{FeatMark} & Y & 0.98 & 0.75 & 0.96 & 0.96 & 0.99 & 0.98 \\
& N & 0.03 & 0.04 & 0.04 & 0.04 & 0.04 & 0.05 \\
\midrule
\multicolumn{8}{c}{\textbf{CelebA-HQ}} \\
\multirow{2}{*}{\shortstack{Stable\\Signature}} & Y & 0.99 & 0.38 & 0.98 & 0.97 & 0.66 & 0.82 \\
& N & 0.46 & 0.42 & 0.49 & 0.48 & 0.49 & 0.46 \\
\cmidrule(r){1-8}
\multirow{2}{*}{HiDDeN} & Y & 0.95 & 0.61 & 0.90 & 0.95 & 0.81 & 0.67 \\
& N & 0.45 & 0.46 & 0.42 & 0.42 & 0.42 & 0.47 \\
\cmidrule(r){1-8}
\multirow{2}{*}{StegaStamp} & Y & 0.95 & 0.58 & 0.91 & 0.98 & 0.90 & 0.90 \\
& N & 0.41 & 0.50 & 0.41 & 0.47 & 0.38 & 0.45 \\
\cmidrule(r){1-8}
\multirow{2}{*}{FeatMark} & Y & 0.92 & 0.31 & 0.90 & 0.89 & 0.94 & 0.88 \\
& N & 0.16 & 0.02 & 0.28 & 0.22 & 0.19 & 0.22 \\

\midrule
\multicolumn{8}{c}{\textbf{WikiArt}} \\
\multirow{2}{*}{\shortstack{Stable\\Signature}} & Y & 0.99 & 0.50 & 0.80 & 0.88 & 0.72 & 0.76 \\
& N & 0.47 & 0.56 & 0.46 & 0.53 & 0.48 & 0.52 \\
\cmidrule(r){1-8}
\multirow{2}{*}{HiDDeN} & Y & 0.99 & 0.52 & 0.96 & 0.98 & 0.78 & 0.62 \\
& N & 0.41 & 0.50 & 0.41 & 0.47 & 0.38 & 0.45 \\
\cmidrule(r){1-8}
\multirow{2}{*}{StegaStamp} & Y & 0.99 & 0.48 & 0.97 & 0.99 & 0.88 & 0.85 \\
& N & 0.44 & 0.45 & 0.45 & 0.44 & 0.41 & 0.43 \\
\cmidrule(r){1-8}
\multirow{2}{*}{FeatMark} & Y & 0.99 & 0.26 & 0.77 & 0.96 & 0.99 & 0.91 \\
& N & 0.01 & 0.02 & 0.01 & 0.01 & 0.01 & 0.02 \\

\bottomrule
\end{tabular}
}
\end{table}

\subsection{Watermark Reading}
\label{method:watermark_extraction}

The watermark reader $\mathcal{R}$ determines whether a suspicious image carries the user’s target message $m$. Unlike pixel-level invisible watermark decoders~\cite{hidden,stablesignature}, our watermark signal is semantic and may vary in its localized realization across images; accordingly, we adopt a feature-based reader that leverages a strong frozen image encoder and trains only a lightweight head.

Concretely, we instantiate $\mathcal{R}$ as a frozen CLIP image encoder~\cite{clip} followed by a trainable fully connected layer that outputs a $|m|$-bit vector. Given a small set of clean images $\mathcal{X}=\{x\}$, their watermarked counterparts $\mathcal{X}^{\text{wm}}=\{x^{\text{wm}}\}$, and the target message $m$, we train $\mathcal{R}$ with:
\begin{equation}
\begin{aligned}
\mathcal{L}_{\text{read}} ={}& \mathbb{E}_{x\sim \mathcal{X}}\big[\mathrm{BCE}(\bar{m},\mathcal{R}(x))\big] \\
&+\mathbb{E}_{x^{\text{wm}}\sim \mathcal{X}^{\text{wm}}}\big[\mathrm{BCE}(m,\mathcal{R}(x^{\text{wm}}))\big],
\end{aligned}
\label{eq:loss_read}
\end{equation}
where $\bar{m}$ denotes the bitwise complement of $m$. During both training and inference, standard augmentations (cropping, rotation, mild noise) and weight decay are applied to improve generalization. After training, verification reduces to checking whether the decoded bits from a suspicious image match $m$ above a fixed threshold, enabling attribution of attacker-generated outputs even when the attacker applies post-processing, provided that the semantic watermark evidence persists through the mimicry pipeline.

\section{Evaluations}
\label{sec:exp_result}
In this section, we conduct the experiments regarding perceptual quality evaluation, the watermark detection accuracy, and the robustness of our watermarking method against removal and purification attacks. Additional adaptive attacks, ablations, and generalization results appear in Appendices~\ref{app:adaptive-details}, \ref{sec:exp:ablation}, and~\ref{app:deployment-evidence}. To ensure clarity, in visualization figures we present four samples per set for each row by default, unless otherwise specified.

\subsection{Experimental Setup}
\label{sec:exp_setup}
We evaluate FeatMark on watermark decoding, perceptual quality, and robustness under purification/removal attacks across VGGFace2, CelebA-HQ, and WikiArt. The implementation uses CLIP-based evidence scoring, text-guided open-world segmentation, and masked diffusion editing; mimicry attacks follow a DreamBooth-style protocol, and metrics include bit accuracy, FID, precision/recall, and coverage/density. We keep the main evaluation focused on the primary results and provide full dataset, implementation, metric, baseline, and user-study details in Appendix~\ref{app:full-exp-setup}.

\begin{table*}[t]
\small
\centering
\begin{minipage}[t]{0.48\linewidth}
\centering
\caption{Bit accuracy change of FeatMark against IMPRESS across all datasets.}
\label{table:results_preprocessing}
\fitwidth{
\begin{tabular}{lcccccc}
\toprule
\multirow{2}{*}{\textbf{Dataset}} 
& \multicolumn{6}{c}{\textbf{Change of Bit Accuracy}} \\
\cmidrule(r){2-7}
& None & Crop & Brigh. & Cont. & JPEG & Comb. \\
\midrule
VGGFace2 & $+$0.00 & $+$0.01 & $+$0.00 & $+$0.00 & $+$0.00 & $+$0.00 \\
CelebA-HQ & $-$0.00 & $-$0.01 & $-$0.00 & $-$0.01 & $-$0.00 & $-$0.00 \\
WikiArt & $-$0.00 & $+$0.00 & $+$0.00 & $-$0.01 & $-$0.00 & $-$0.00 \\
\bottomrule
\end{tabular}}
\end{minipage}
\hfill
\begin{minipage}[t]{0.48\linewidth}
\centering
\caption{Bit accuracy change of FeatMark against Noisy Upscaling across all datasets.}
\label{table:results_upscaling}
\fitwidth{
\begin{tabular}{lcccccc}
\toprule
\multirow{2}{*}{\textbf{Dataset}} 
& \multicolumn{6}{c}{\textbf{Change of Bit Accuracy}} \\
\cmidrule(r){2-7}
& None & Crop & Brigh. & Cont. & JPEG & Comb. \\
\midrule
VGGFace2 & $-$0.20 & $-$0.01 & $-$0.26 & $-$0.24 & $-$0.31 & $-$0.26 \\
CelebA-HQ & $-$0.22 & $-$0.02 & $-$0.24 & $-$0.19 & $-$0.26 & $-$0.30 \\
WikiArt & $-$0.23 & $-$0.01 & $-$0.27 & $-$0.21 & $-$0.29 & $-$0.25 \\
\bottomrule
\end{tabular}}
\end{minipage}
\end{table*}

\begin{figure*}[t]
\centering
\includegraphics[width=1\linewidth]{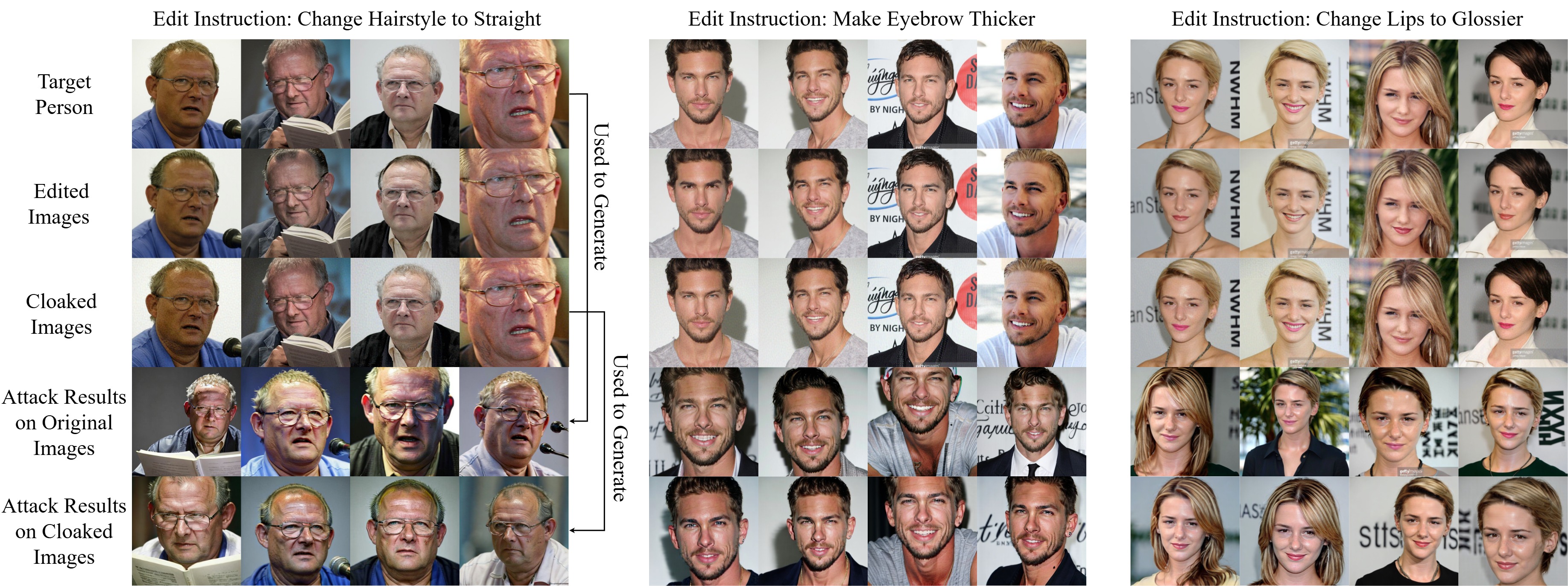}
\caption{\textbf{FeatMark on VGGFace2.} Three examples with the edit instruction shown on top. The left example annotates the row order, reused for the other two. Rows show target images, edited images, cloaked released images, and attacker outputs trained on original images and on cloaked images. Training on cloaked images makes the injected attribute consistently appear in the attacker generations.}
\label{fig:visualize_vggface}
\Description{} \end{figure*}

\begin{table*}[t]
\small
\centering
\begin{minipage}[t]{0.46\linewidth}
\centering
\caption{Bit accuracy changes of three watermarking schemes in VGGFace2 against three watermark removal attacks, UnMarker, Diffusion Attack, and VAE Attack.}
\label{table:results_unmarker}
\fitwidth{
\begin{tabular}{lccc}
\toprule
& UnMarker & Diffusion Attack & VAE Attack \\
\midrule 
Stable Signature & $-$0.99 & $-$0.99 & $-$0.99 \\
Tree-Ring & $-$0.39 & $-$0.21 & $-$0.09 \\
FeatMark & $-$0.01 & $-$0.01 & $-$0.00 \\
\bottomrule
\end{tabular}
}
\end{minipage}\hfill
\begin{minipage}[t]{0.50\linewidth}
\centering
\caption{Bit accuracy changes of four watermarking schemes in VGGFace2 against four watermark removal attacks in Watermark Attacker (WMA).}
\label{table:results_watermark_attacker}
\fitwidth{
\begin{tabular}{lcccc}
\toprule
& WMA-Diff60 & WMA-Cheng3 & WMA-BMSHJ3 & WMA-JPEG50 \\
\midrule 
DwtDct & $-$0.33 & $-$0.33 & $-$0.30 & $-$0.29 \\
DwtDctSvd & $-$0.45 & $-$0.47 & $-$0.43 & $-$0.20 \\
RivaGAN & $-$0.45 & $-$0.35 & $-$0.39 & $-$0.02 \\
FeatMark & $-$0.02 & $-$0.00 & $-$0.00 & $-$0.01 \\
\bottomrule
\end{tabular}
}
\end{minipage}
\end{table*}

\begin{figure*}[t]
\centering
\includegraphics[width=1\linewidth]{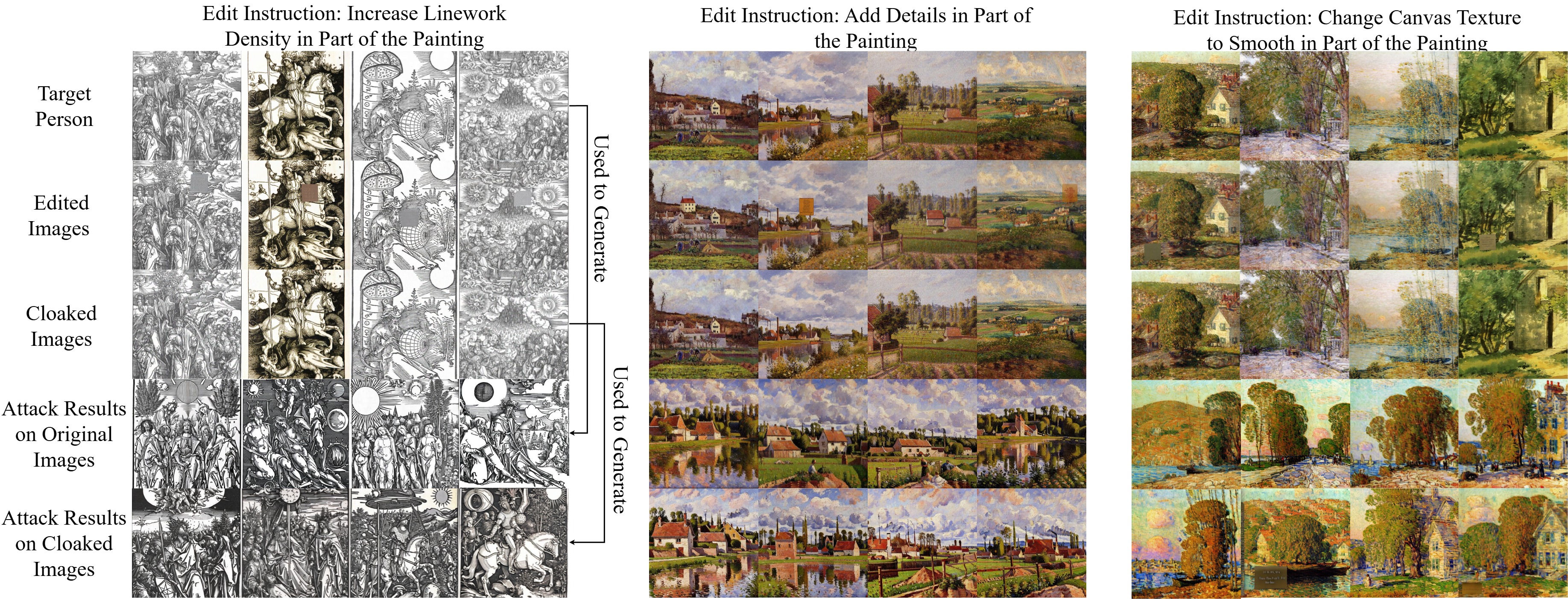}
\caption{\textbf{FeatMark on WikiArt.} Three examples with the edit instruction shown on top. The left example annotates the row order, reused for the other two. Rows show target artworks, edited artworks, cloaked released artworks, and attacker outputs trained on original artworks and on cloaked artworks. Models trained on cloaked artworks inherit the injected local style change.}
\label{fig:visualize_wikiart}
\Description{} \end{figure*}

\subsection{Perceptual Quality Evaluation}
\label{sec:perceptual_quality}
\label{app:qualitative}

Perceptual quality of FeatMark images is measured in terms of the average FID, precision, recall, coverage, and density for all target persons. 
These are comparative metrics between a source and target data sample, as such for each target identity, the generated fake images impacted by our watermark is contrasted against the original images of that identity.  
As the name suggests, No-Defense is calculated with fake images that had no defense applied.

\noindent \textbf{Perceptual results.~}
Table~\ref{table:results_perceptual_all} shows that FeatMark preserves substantially higher perceptual quality than disruption-based defenses across all datasets. Compared to the No-Defense setting, FeatMark induces only a modest degradation in distributional fidelity: the FID increases by $9.17$ on VGGFace2 ($180.98$ vs.\ $171.81$), $13.34$ on CelebA-HQ ($142.33$ vs.\ $128.99$), and $15.74$ on WikiArt ($321.76$ vs.\ $306.02$). Importantly, FeatMark maintains non-trivial Precision/Recall as well as Coverage/Density, indicating that the generated samples remain both realistic and diverse under our fully automatic semantic watermark injection.

In contrast, existing image cloaking methods severely damage perceptual quality. For examples, FIDs for Anti-Dreambooth and Glaze balloon to $449.74$--$465.37$ on VGGFace2/CelebA-HQ and $410.93$ on WikiArt, while Precision, Recall, Coverage, and Density collapse to zero. This stark gap highlights FeatMark’s key practical advantage: it enables reliable downstream attribution while keeping the attacker’s outputs close to the original data distribution, rather than relying on heavy distortions that break visual fidelity.

\noindent \textbf{Visualization of FeatMarks.~}
Figures~\ref{fig:visualize_vggface} and~\ref{fig:visualize_wikiart} show VGGFace2 and WikiArt results; CelebA-HQ examples appear in Appendix~\ref{app:additional-qualitative}. Across VGGFace2, CelebA-HQ, and WikiArt, the visual examples show the same pattern: released cloaked images remain close to the targets, while attackers trained on cloaked images consistently inherit the injected semantic attribute.

\noindent \textbf{User study results.~}
\label{app:user-study-results}
We compare 100 pairs of attacker-generated images from No-Defense and FeatMark using 49 participants. Each pair uses the same target and prompt, with randomized left--right ordering. Participants judge realism, identity/style match, and agreement with the edit instruction. The study protocol is detailed in Appendix~\ref{app:full-exp-setup}.

\begin{table}[t]
\small
\setlength{\tabcolsep}{3pt}
\centering
\caption{User study on attacker-generated images under No-Defense vs FeatMark.
Each entry is the FeatMark win rate in a two-alternative forced choice test with randomized left-right ordering.
For realism and identity or style match, results closer to 50\% indicate near-random guessing and therefore higher perceptual indistinguishability.
For instruction match, higher win rate indicates stronger and more consistent concept transfer.}
\label{tab:user_study_gen_only}
\fitwidth{
\begin{tabular}{lccc}
\toprule
\textbf{Dataset} &
\textbf{Realism win rate} &
\thead{Identity or style\\match win rate} &
\thead{Instruction\\match win rate} \\
\midrule
VGGFace2   & $47.8\% \pm 2.6$ & $50.6\% \pm 2.5$ & $83.1\% \pm 1.9$ \\
CelebA-HQ  & $49.2\% \pm 2.5$ & $51.3\% \pm 2.4$ & $80.4\% \pm 2.1$ \\
WikiArt    & $48.5\% \pm 2.7$ & $49.8\% \pm 2.6$ & $77.6\% \pm 2.4$ \\
\bottomrule
\end{tabular}}
\end{table}
Table~\ref{tab:user_study_gen_only} shows that FeatMark remains close to the No-Defense baseline for realism and identity/style match, with win rates near 50\%, while strongly improving instruction match. This supports the intended tradeoff: the generated images remain perceptually comparable to the baseline, but the injected concept is reliably absorbed and reproduced during mimicry fine-tuning.


\subsection{Watermark Accuracy Evaluation}
\label{sec:watermark_accuracy}

Table~\ref{table:results_bit_acc_all} reports bit accuracy under common transformations. FeatMark maintains high decoding reliability on watermarked images and a large verification margin against non-watermarked images: on VGGFace2 it reaches 0.99/0.98 under JPEG/combined transformations while keeping WM=N accuracy at 0.04/0.05, and on WikiArt it reaches 0.99/0.91 while keeping WM=N accuracy at 0.01/0.02. Cropping remains the most challenging case because it can remove the edited region, but FeatMark still preserves substantially cleaner separability than pixel-level baselines; additional discussion is provided in Appendix~\ref{app:watermark-accuracy-discussion}.


\subsection{Watermark Robustness against Adversarial Purification and Watermark Removal}
\label{sec:watermark_robustness}
\label{app:robustness-details}

We stress-test FeatMark against adversarial purification, white-box perturbation attacks, and black-box removal/regeneration attacks. Tables~\ref{table:results_preprocessing}--\ref{table:results_upscaling} summarize the purification results; Tables~\ref{table:results_unmarker} and~\ref{table:results_watermark_attacker} report black-box removal results, with the Noisy Upscaling visualization in Appendix~\ref{app:additional-qualitative}. Attack configurations are described in Appendix~\ref{app:full-exp-setup}.

\noindent \textbf{Purification attacks.~}
IMPRESS causes essentially no bit-accuracy change across datasets, while Noisy Upscaling produces a measurable but non-decisive reduction of about $0.20$--$0.31$ under most transformations. Even after Noisy Upscaling, FeatMark remains decodable from its high baseline accuracy, and the reduction largely coincides with visible rewriting artifacts.

\noindent \textbf{Removal and regeneration attacks.~}
WEvade achieves evasion rates of $1.00$ for Stable Signature but only $0.16$ for FeatMark in the removal setting, and $1.00$ vs. $0.14$ in the forgery setting. Black-box removal and regeneration attacks also have little effect on FeatMark: UnMarker, diffusion regeneration, VAE reconstruction, and WatermarkAttacker change FeatMark's bit accuracy by at most about $0.02$, while substantially degrading pixel- or coefficient-level baselines.

\noindent \textbf{Video mimicry and adaptive attacks.~}
FeatMark also extends to video mimicry, where the verifier achieves $0.85$ bit accuracy on watermarked videos versus $0.13$ on non-watermarked videos. Details are provided in Appendix~\ref{app:video-mimicry}. We further evaluate mechanism-aware adaptive attacks in Appendix~\ref{app:adaptive-details}.

\section{Conclusion}
We presented FeatMark, a feature-level watermarking framework for attributing diffusion-based mimicry attacks. FeatMark encodes provenance as inconspicuous semantic micro-features selected automatically from a feature bank and transferred through mimicry training via image cloaking. Experiments on faces, artworks, and video show near--No-Defense perceptual quality, clear watermarked/non-watermarked separation, and strong robustness to purification, removal, and tailored adaptive attacks.

{\small
\bibliographystyle{ieeenat_fullname}
\bibliography{sample-base}
}


\clearpage
\appendix
\section*{Appendix}

\section{Full Experimental Setup}
\label{app:full-exp-setup}
We evaluate FeatMark from three perspectives: (1) \emph{watermark detection accuracy} on attacker-generated images, (2) \emph{perceptual quality} of the generated outputs, and (3) \emph{robustness} of watermark verification under purification and watermark-removal attacks. In every experiment, the attacker receives one of three training sets, original images (No-Defense), images protected by a baseline watermark/defense, or FeatMark-protected released images, and performs mimicry fine-tuning to generate fake images. We then evaluate those fakes using the metrics below.

\noindent\textbf{Datasets.~}
We follow the dataset protocols used in prior cloaking and purification evaluations~\cite{anti-dreambooth, glaze, impress}. We consider two face datasets (VGGFace2~\cite{vggface2} and CelebA-HQ~\cite{celebahq1, celebahq2}) and one artwork dataset (WikiArt~\cite{wikiart}). We use the standard partitions and mimicry settings from DreamBooth-style evaluations~\cite{dreambooth, anti-dreambooth, impress}. Since VGGFace2 has higher identity/attribute diversity, we additionally use it for adaptive-attack robustness and ablations~\cite{anti-dreambooth, vggface2}.

\noindent\textbf{Implementations.~}
We instantiate the open-vocabulary evidence scorer with CLIP~\cite{clip}. For region grounding $M_k(x)$ we use a text-guided open-world segmentation pipeline (LangSAM) built upon GroundingDINO~\cite{liu2024groundingdino} and Segment Anything~\cite{kirillov2023segment} via Grounded-SAM~\cite{ren2024groundedsam}. For the concept-guided editor $\mathcal{E}$ in Eq.~\eqref{eq:edit}, we use a diffusion-based image-to-image inpainting pipeline conditioned on the edit instruction and mask; other instruction-following masked editors can be substituted without changing the methodology.

\noindent\textbf{Training configuration.~}
We optimize each image cloak for 500 iterations with step size $1.0/255$, $\lambda=10^2$ in Eq.~\eqref{eq:loss_cloak}, and default perturbation budget $0.10$ (ablations in Appendix~\ref{sec:exp:ablation}). Mimicry attacks use DreamBooth~\cite{dreambooth} for 1000 steps on a Stable Diffusion v2.1 base~\cite{stable_diffusion} (additional diffusion versions and personalization backends are evaluated in Appendices~\ref{sec:diffusion-generalization} and~\ref{app:personalization-backends}). The watermark extractor is trained for 100 iterations with AdamW ($lr=10^{-2}$, $weight\_decay=10^{-2}$); we use 48-bit messages and follow HiDDeN~\cite{hidden} to train the output layer with AdamW ($lr=0.01$, $weight\_decay=0.01$)~\cite{adamw}. Unless otherwise stated, attribution is evaluated by bit accuracy and a fixed decision threshold chosen on a held-out calibration split; all WM=Y/WM=N tables report the resulting separation rather than a formal proof of verification.

\noindent\textbf{Compute resources.~}
All experiments were conducted on NVIDIA A100X GPUs with 80GB memory, using
NVIDIA driver 550.54.14 and CUDA 12.4. 
A DreamBooth mimicry run with
1,000 fine-tuning steps took about 10 minutes on one A100X GPU. LoRA and Textual
Inversion runs were lighter and took approximately 3--6 minutes per target. The
remaining stages, including concept selection, localized editing, image cloaking,
reader training, watermark extraction, purification/removal attacks, and adaptive
attack evaluation, were run on the same GPU type. The complete
experimental suite consumed approximately 150--250 A100X GPU-hours in total.

\noindent\textbf{Evaluation metrics.~}
Watermark accuracy is measured by bit accuracy (fraction of correctly decoded bits). Unless stated otherwise, we report robustness under common transformations~\cite{stablesignature}: strong crop (retain 90\%), brightness $\times 2.0$, contrast $\times 2.0$, JPEG quality 50, and a combined transform (Crop+Brigh.+JPEG). Perceptual quality is evaluated on the \emph{attacker-generated} images using FID~\cite{fid}, precision/recall~\cite{precision_recall}, and coverage/density~\cite{coverage_density} between generated fakes and the original images.

\noindent\textbf{Removal and purification attacks.~}
To stress-test robustness, we evaluate purification attacks applied before mimicry training (IMPRESS~\cite{impress} and Noisy Upscaling~\cite{adversarial_perturbations_cannot, upscale}), a white-box watermark remover with extractor access (WEvade~\cite{wevade}), and black-box removal/regeneration attacks including UnMarker~\cite{unmarker}, diffusion-based regeneration~\cite{diffpure}, VAE reconstruction~\cite{vae}, and the WatermarkAttacker framework with multiple configurations~\cite{watermark_attacker}.

\noindent\textbf{Baselines.~}
We compare against representative invisible watermark baselines for diffusion models (Stable Signature~\cite{stablesignature} and Tree-Ring~\cite{tree-ring-watermarks}). 
We additionally include learned message watermarking baselines HiDDeN~\cite{hidden} and StegaStamp~\cite{stegastamp}. Since prior image cloaks are not designed for verification, we also include Anti-DreamBooth~\cite{anti-dreambooth}, and for WikiArt we include Glaze~\cite{glaze}, to contextualize perceptual quality. We further add MetaCloak~\cite{metacloak}, a meta-learning-based poisoning defense designed to impede subject-driven diffusion fine-tuning. 
For removal-attack comparisons, we additionally report classic frequency-/learning-based schemes (DwtDct, DwtDctSvd, RivaGAN) as used by WatermarkAttacker~\cite{watermark_attacker}.

\noindent\textbf{User study.~}
We conduct a pairwise study comparing attacker-generated images from No-Defense versus FeatMark, using the same target and prompt per pair. Order is randomized and balanced. Participants answer which image is more realistic, which better matches the target identity/style, and which better matches the edit instruction, following standard pair-comparison guidance~\cite{ituBT500,ituP915,mantiuk2012comparison}. We recruit 49 participants on Prolific~\cite{prolific} to label 100 image pairs (18+, no special face-recognition expertise), with a 30-minute limit and \pounds 3.00 compensation.

\section{Additional Watermark Accuracy Discussion}
\label{app:watermark-accuracy-discussion}
A key advantage of FeatMark is the verification margin between WM=Y and WM=N, which directly translates into lower false positives and more reliable attribution. For VGGFace2 and WikiArt, FeatMark drives WM=N accuracy close to zero across all transformations (e.g., 0.03 to 0.05 on VGGFace2 and 0.01 to 0.02 on WikiArt), while keeping WM=Y accuracy high. Even under cropping, FeatMark preserves strong separability, with 0.75 vs.\ 0.04 on VGGFace2 and 0.26 vs.\ 0.02 on WikiArt. In contrast, Stable Signature exhibits WM=N accuracies near random guessing around 0.45--0.56, and most critically its separability collapses under cropping where WM=Y becomes comparable to WM=N (e.g., 0.36 vs.\ 0.40 on VGGFace2, 0.38 vs.\ 0.42 on CelebA-HQ, and 0.50 vs.\ 0.56 on WikiArt). While all these methods can decode well in benign conditions, FeatMark more consistently maintains a usable decision margin when the attacker applies realistic removal attack. Cropping is inherently the most challenging transformation for localized semantic watermarks, since it can remove the edited region entirely. Nevertheless, FeatMark’s design still yields a practical attribution signal whenever part of the marked region remains, and its strong suppression of WM=N decoding provides a cleaner and more reliable verification behavior than baselines across datasets and perturbations.

\section{Additional Qualitative Results}
\label{app:additional-qualitative}
The CelebA-HQ examples complement the VGGFace2 and WikiArt visualizations in the main paper. We also provide the Noisy Upscaling visualization for the purification results discussed in Section~\ref{sec:watermark_robustness}.

\begin{figure*}[t]
\centering
\includegraphics[width=1\linewidth]{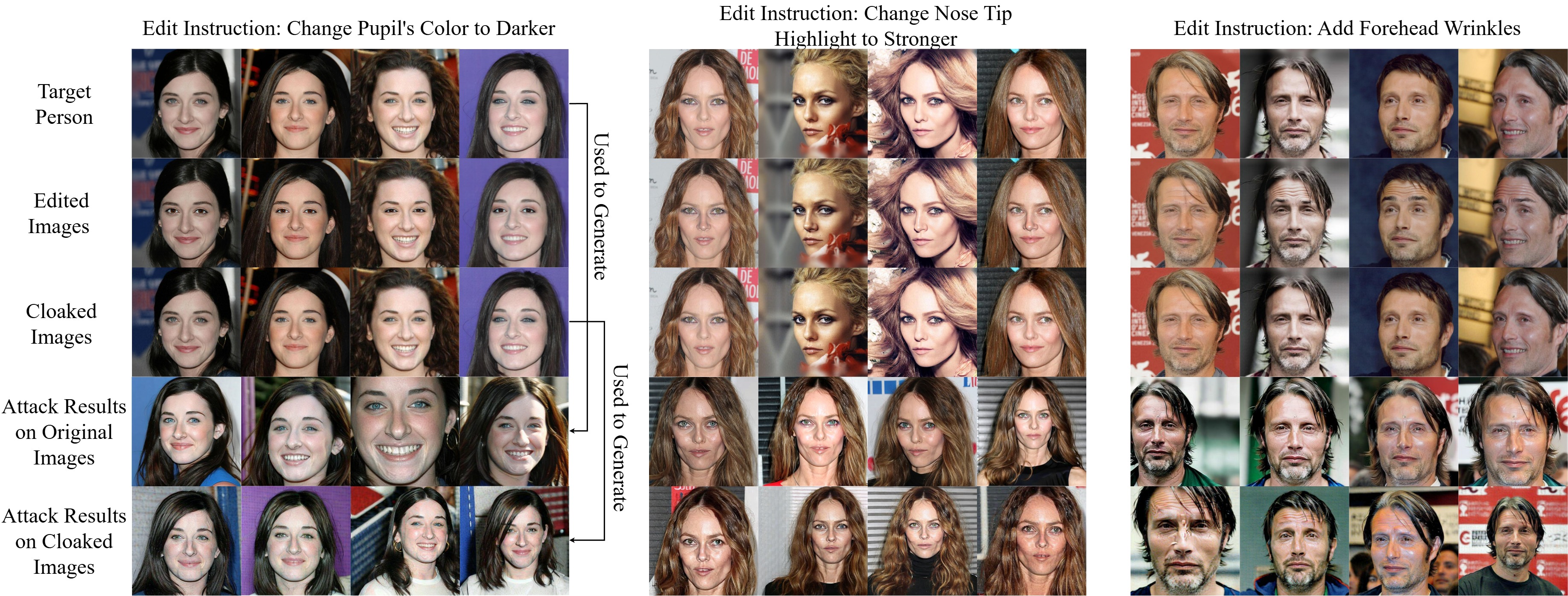}
\caption{\textbf{FeatMark on CelebA-HQ.} Three examples with the edit instruction shown on top. The left example annotates the row order, reused for the other two. Rows show target images, edited images, cloaked released images, and attacker outputs trained on original images and on cloaked images. The injected facial concept transfers to the attacker generations when trained on cloaked images.}
\label{fig:visualize_celebahq}
\Description{} \end{figure*}

\begin{figure}[t]
\centering
\includegraphics[width=\linewidth]{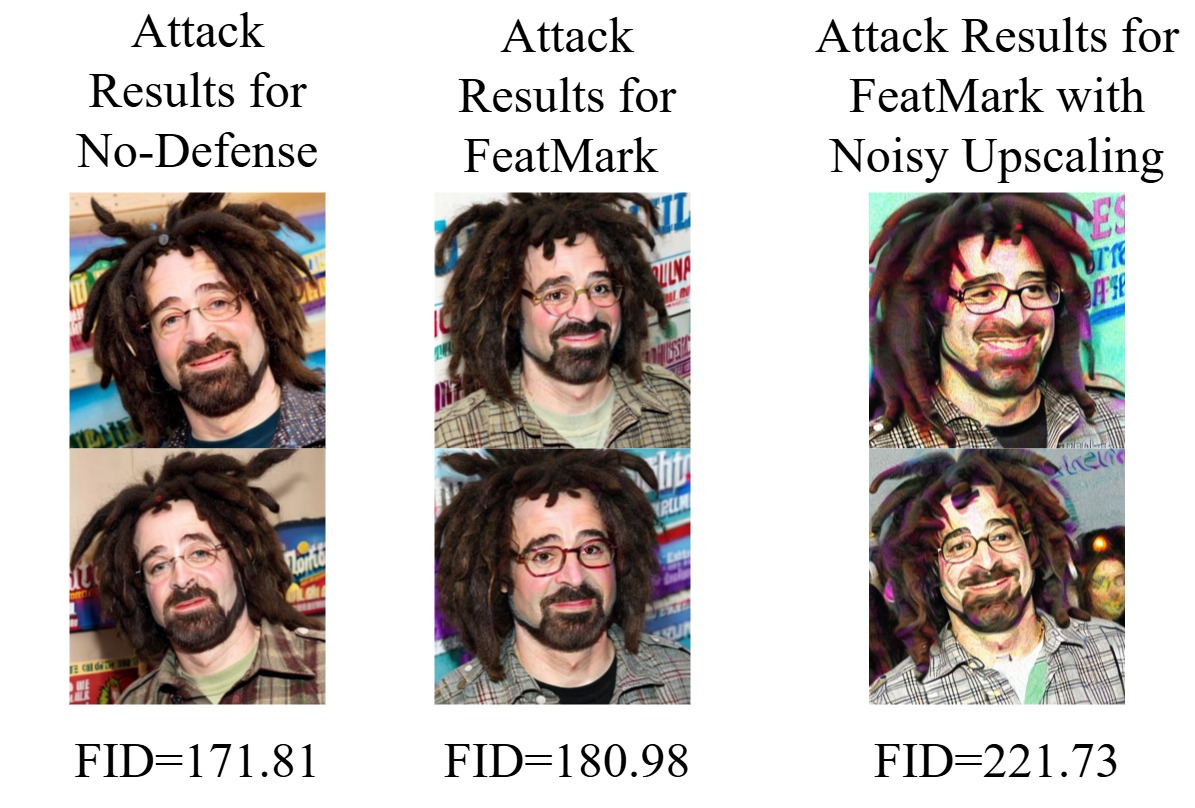}
\caption{Visualization of FeatMark against Noisy Upscaling for one target person in VGGFace2 dataset.}
\label{fig:visualize_n000068_upscale}
\Description{}
\end{figure}

\section{Video Mimicry Extension}
\label{app:video-mimicry}
\subsection{Extensibility to Video Mimicry}
\label{sec:video-mimicry}

We extend FeatMark from images to the video mimicry setting by replacing the image editor with a masked video-to-video editor. Concretely, for each target video we first select one subtle feature from our feature bank, then localize it on the first frame via text-guided segmentation and propagate the mask across the entire clip using a promptable video segmentation tracker (SAM~2)~\cite{ravi2024sam2}. Given the propagated mask sequence, we apply a single concept instruction consistently to all frames via a masked video editing backbone, yielding an edited video whose watermark evidence is temporally coherent.We then adapt our cloaking stage from a single image to a video by distributing the cloaking perturbation across frames. Specifically, we add bounded, per-frame noise after editing so that the published video remains visually faithful while still steering the downstream mimicry process toward reproducing the injected feature evidence throughout the clip.

On the attacker side, we instantiate video mimicry using CogVideoX~\cite{yang2024cogvideox} and perform LoRA-based~\cite{hu2021lora} fine-tuning on the target character videos from CREMA-D~\cite{cao2014cremad}. Evaluating the resulting fake videos, our verifier achieves a bit accuracy of $0.13$ on non-watermarked samples and $0.85$ on watermarked samples, indicating a clear separation between clean and FeatMark-protected content in the video domain.


\section{Adaptive Attack Details}
\label{app:adaptive-details}
The main paper summarizes the adaptive attacks and keeps the full quantitative table and qualitative visualization here.

\begin{table}[t]
\small
\centering
\caption{Bit accuracy change of FeatMark against tailored adaptive attacks on VGGFace2.}
\label{table:results_adaptive}
\fitwidth{
\begin{tabular}{lcccccc}
\toprule
\multirow{2}{*}{\textbf{Attack}} 
& \multicolumn{6}{c}{\textbf{Change of Bit Accuracy}} \\
\cmidrule(r){2-7}
& None & Crop & Brigh. & Cont. & JPEG & Comb. \\
\midrule
TCN & $-$0.02 & $-$0.00 & $-$0.07 & $-$0.01 & $-$0.02 & $-$0.09 \\
AFT & $-$0.01 & $-$0.00 & $-$0.03 & $-$0.01 & $-$0.02 & $-$0.01 \\
DRA-G & $-$0.20 & $-$0.02 & $-$0.22 & $-$0.26 & $-$0.20 & $-$0.23 \\
DRA-G & $-$0.22 & $-$0.07 & $-$0.25 & $-$0.27 & $-$0.24 & $-$0.32 \\
\bottomrule
\end{tabular}
}
\end{table}

\subsection{Attack Methodology}
\label{sec:adaptive-attack-suite}

We evaluate three adaptive attackers with increasing capability, and refer to them by short names in later results.

\textbf{TCN (Targeted Concept Neutralization).~}
FeatMark’s watermark is realized as a localized concept, so a natural adaptive strategy is to infer the concept and undo it. TCN ranks candidate bank features using open-vocabulary concept evidence, localizes the predicted feature region with text-guided grounding, and applies a counterfactual edit that neutralizes the attribute while preserving identity and global semantics. We instantiate grounding with open-world detection and segmentation tools commonly assembled as Grounded-SAM~\cite{liu2024groundingdino,kirillov2023segment,ren2024groundedsam}. This attack is practical because it uses off-the-shelf grounding and editing modules and is directly aligned with the same ``detect--ground--edit'' abstraction as our concept programs, making it a targeted stress test for feature-anchored watermarks.

\textbf{AFT (Adversarial Fine-Tuning).~}
A stronger attacker can push watermark evasion into training, so that the generator learns to avoid verifiable outputs by construction. Assuming white-box access to the reader $\mathcal{R}$ and message $m$, the attacker fine-tunes mimicry model parameters $\theta$ to suppress decoded bits while preserving identity or style relative to a reference checkpoint $\theta_0$. Using a deterministic sampling routine $x_{\theta}(z,c)$ from the mimicry model for noise seed $z$ and condition $c$, a representative objective is
\begin{equation}
\begin{aligned}
\min_{\theta}\quad &\mathbb{E}_{z,c}\Big[\mathrm{BCE}\big(m,\mathcal{R}(x_{\theta}(z,c))\big)\Big] \\
\text{s.t.}\quad &\mathrm{Sim}\big(x_{\theta}(z,c),x_{\theta_0}(z,c)\big) \ge \gamma.
\end{aligned}
\label{eq:aft}
\end{equation}
As such, AFT provides a meaningful upper bound because it directly optimizes against the exact verification mechanism under an explicit fidelity constraint.

\textbf{DRA (Diffusion Regeneration Attack).~}
DRA models a practical ``regenerate-to-remove'' threat where an attacker applies diffusion-based image-to-image regeneration to rewrite details while keeping global content largely intact, as in SDEdit-style editing and diffusion purification~\cite{meng2022sdedit}. We evaluate two variants. \textbf{DRA-UG} (unguided) simply regenerates each image with a tunable strength (e.g., noise level or diffusion steps) and selects the strongest setting that stays within a perceptual similarity budget. \textbf{DRA-G} (guided) additionally uses the watermark reader as feedback, choosing regeneration settings---and optionally adding reader-guidance during regeneration---to directly minimize verification, e.g., by reducing $\mathcal{B}_{\mathrm{acc}}\!\left(m,\mathcal{R}(\tilde{x})\right)$ (or an equivalent differentiable surrogate such as $\mathrm{BCE}(m,\mathcal{R}(\tilde{x}))$). Together, DRA-UG and DRA-G capture both an off-the-shelf post-processing attacker and a stronger mechanism-aware variant that explicitly targets the decoded bits.

\begin{figure}[t]
\centering
\includegraphics[width=\linewidth]{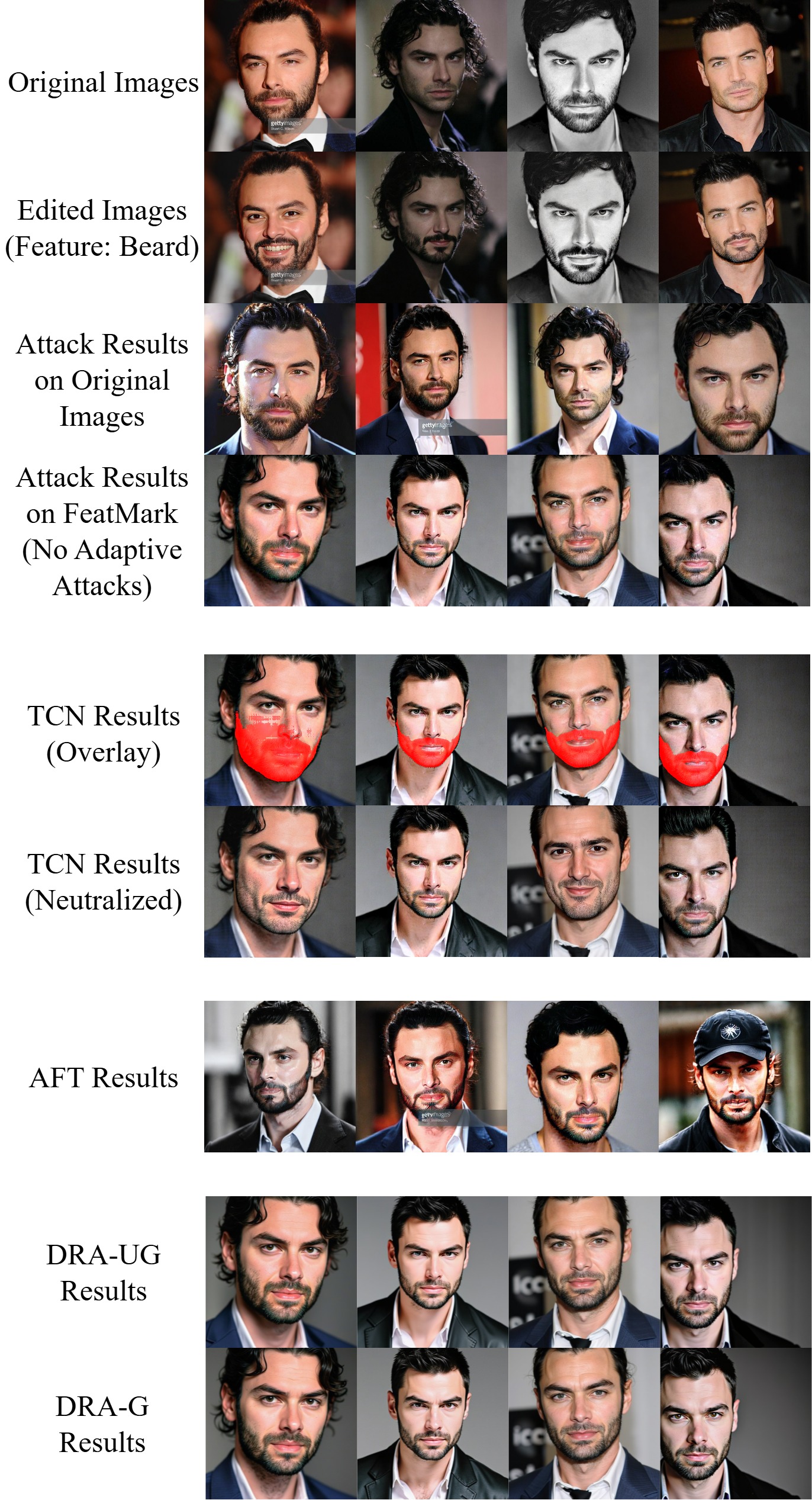}
\caption{Visualization of FeatMark against three adaptive attacks for one target person in VGGFace2 dataset.}
\label{fig:visualize_adaptive}
\Description{}
\end{figure}

\section{Related Work}
\label{app:related-work}
\label{sec:related}

We position FeatMark against four lines of work: personalization attacks, watermark-based attribution, anti-mimicry cloaks, and watermark/cloak removal attacks. This organization is important because many defenses that look superficially comparable optimize for a different security goal.

\noindent \textbf{Personalization and mimicry attacks.~}
Modern mimicry pipelines personalize text-to-image generators from a small reference set. DreamBooth binds a rare token to a subject or style through fine-tuning~\cite{dreambooth}, Textual Inversion learns a new textual embedding while keeping the generator mostly fixed~\cite{textual_inversion}, and parameter-efficient variants such as LoRA update low-rank adapters rather than all model weights~\cite{hu2021lora}. Other personalization interfaces, such as Custom Diffusion~\cite{custom_diffusion} and IP-Adapter~\cite{ip_adapter}, further broaden the attacker design space by changing how image identity/style is injected. FeatMark is not tied to a single personalization implementation: its signal is placed in localized semantic evidence that a successful mimicry model must reproduce to imitate the target. Appendix~\ref{app:personalization-backends} therefore evaluates LoRA and Textual Inversion in addition to the DreamBooth protocol used in the main experiments.

\noindent \textbf{Watermarks and provenance.~}
Generation-time watermarks integrate provenance into the synthesis process. Stable Signature~\cite{stablesignature}, Tree-Ring~\cite{tree-ring-watermarks}, Gaussian Shading~\cite{gaussian_shading}, ZoDiac~\cite{zodiac}, WaDiff~\cite{a_watermark_conditioned_dm}, EditGuard~\cite{editguard}, and WMCopier~\cite{wmcopier} embed or propagate low-level signals that can be decoded later; earlier image watermarking systems such as HiDDeN~\cite{hidden} and StegaStamp~\cite{stegastamp} provide learned message decoders for natural images. Source-side residue methods place an imperceptible mark on the public training images before release, including ProMark~\cite{promark}, GenWatermark~\cite{genwatermark}, DIAGNOSIS~\cite{diagnosis}, FT-Shield~\cite{ft-shield}, and related concept-watermarking schemes for textual inversion~\cite{guarding_textual_inversion}. Metadata standards such as C2PA~\cite{c2pa_spec} are complementary but do not survive the training process once an adversary extracts raw pixels. FeatMark differs in mechanism and threat model: it makes the provenance signal part of the visual/semantic data distribution that the mimicry model learns, rather than relying on a low-energy pixel signature or a file-container credential.

\noindent \textbf{Anti-mimicry cloaks and style-protection defenses.~}
Cloaking defenses add adversarial perturbations to public images so that personalization learns the wrong identity/style. Representative methods include AdvDM~\cite{advdm}, PhotoGuard~\cite{photoguard}, Anti-DreamBooth~\cite{anti-dreambooth}, Glaze~\cite{glaze}, MetaCloak~\cite{metacloak}, Nightshade~\cite{nightshade}, LightShed~\cite{lightshed}, and related image-cloak analyses~\cite{effective_image_cloak,adversarial_perturbations_cannot}. Recent style-specific defenses such as StyleGuard~\cite{styleguard} and StyleProtect~\cite{styleprotect} strengthen perturbation-based style protection by targeting style-related features or selected diffusion components. These works are important baselines for \emph{disruption}: they try to make the attacker fail to reproduce the original identity/style. FeatMark instead targets \emph{attribution}: it intentionally preserves usable mimicry quality while causing successful mimicry outputs to carry a decodable semantic trace. This distinction explains why disruption metrics alone are insufficient for evaluating FeatMark and why our main tables report WM=Y/WM=N separability, bit accuracy, and perceptual quality rather than only attack failure rates.

\noindent \textbf{Removal, purification, and adaptive trigger-inference attacks.~}
Pixel-level watermarks and cloaks can be attacked through adversarial perturbation, regeneration, or spectral manipulation. WEvade~\cite{wevade,guo2024aigeneratedimagedetectionpassive}, surrogate-key optimization~\cite{adaptive_attacks_watermarks}, watermark forgery~\cite{warfare}, unstable diffusion-component fine-tuning~\cite{stable_signature_is_unstable}, transfer attacks~\cite{transfer_attack_image_watermarks}, UnMarker~\cite{unmarker}, WatermarkAttacker~\cite{watermark_attacker}, and latent VAE attacks~\cite{a_crack_in_the_bark} demonstrate that invisible or coefficient-level signals are often fragile. Reviewers may also ask whether trigger-recovery methods such as Neural Cleanse~\cite{neural_cleanse}, originally designed for backdoor triggers, can reverse-engineer FeatMark. We address this in Appendix~\ref{app:trigger-inversion}: FeatMark is not a single universal patch or fixed trigger. The selected cue, localization, edit direction/strength, and cloaking objective are sample-conditioned and defender-side, so trigger inversion can at best infer a coarse semantic hypothesis; removing the transferred evidence generally requires visible identity/style edits rather than benign denoising.

\section*{Declaration of LLM Usage}
A generative AI assistant was used only for grammar and light style polishing. It was not used as an original, non-standard, or decision-making component of the research method, experiments, figures, or analysis.


\section*{Ethical Considerations}

This work aims to mitigate harms from diffusion-based mimicry attacks, which can violate personal autonomy, privacy, and reputation by enabling convincing forgeries of an individual’s identity or an artist’s style. FeatMark is designed as a defensive, post-hoc attribution mechanism. Rather than improving synthesis quality or facilitating impersonation, it enables reliable verification of whether suspicious content is likely produced by a mimicry-trained model.

We follow the Menlo Report principles for ICT research~\cite{menlo_report}. Our primary stakeholders include affected individuals and artists whose publicly shared images can be misused for impersonation, researchers and practitioners who build or audit generative systems, and companies/platforms that host or deploy personalization tools. The anticipated benefits are improved provenance and accountability for misuse. Potential risks include that parts of our evaluation pipeline (e.g., mimicry fine-tuning scripts) could be repurposed to lower the barrier for abuse. The implementation and attack configurations are not publicly released. We also do not release any additional identity-specific data beyond what is already publicly available in the referenced datasets, and our verifier parameters remain private by design.

All experiments are conducted offline on publicly available datasets (VGGFace2, CelebA-HQ, WikiArt) following prior protocols, and no live systems, services, or user accounts are targeted or affected. Our user-perception study was reviewed by the authors’ institutional Human Research Ethics Committee (IRB) and determined to be exempt (minimal risk). Participants provided informed consent, were told the study purpose at a high level, and no personally identifying information was collected; responses were used solely for academic research.

Overall, we believe the societal benefit of enabling attribution and accountability for mimicry-generated forgeries outweighs the residual misuse risk, and we explicitly incorporate mitigations through stakeholder analysis and restrictions on artifact distribution.

\section{Additional Generalization and Deployment Evidence}
\label{app:deployment-evidence}

\subsection{Generalization across Diffusion Models}
\label{sec:diffusion-generalization}
Table~\ref{table:results_diffv} evaluates transfer when the attacker uses a different diffusion backbone from the defender. We keep the watermarking model at Stable Diffusion v2.1 and vary the mimicry model across v2.0, v1.5, and v1.4. Without transformations, the corresponding WM=Y/WM=N bit accuracies are $0.97/0.07$, $0.96/0.05$, and $0.98/0.04$, respectively, compared with $0.98/0.03$ for the default v2.1. These results preserve a large verification margin across the tested backbones, supporting attribution without requiring the attacker to use the defender's model version.

\begin{table}[t]
\small
\centering
\caption{Bit accuracy of FeatMark when mimicry attacks are conducted with different versions of diffusion model. v2.1 is the default setting for our watermarks and mimicry attacks (*).}
\label{table:results_diffv}
\fitwidth{
\begin{tabular}{ccl@{\hspace{5pt}}l@{\hspace{5pt}}l@{\hspace{5pt}}l@{\hspace{5pt}}l@{\hspace{5pt}}l}
\toprule
\multirow{2}{*}{\thead{Model\\version}} & \multirow{2}{*}{\textbf{WM?}} & \multicolumn{6}{c}{\textbf{Bit Accuracy}} \\
\cmidrule(r){3-8}
& & None & Crop & Brigh. & Cont. & JPEG & Comb. \\
\midrule

\multirow{2}{*}{v2.1*} & Y & 0.98 & 0.75 & 0.96 & 0.96 & 0.99 & 0.98 \\
& N & 0.03 & 0.04 & 0.04 & 0.04 & 0.04 & 0.05 \\
\midrule
\multirow{2}{*}{v2.0} & Y & 0.97 & 0.72 & 0.94 & 0.97 & 0.98 & 0.99 \\
& N & 0.07 & 0.03 & 0.09 & 0.13 & 0.11 & 0.15 \\
\midrule
\multirow{2}{*}{v1.5} & Y & 0.96 & 0.70 & 0.96 & 0.96 & 0.97 & 0.97 \\
& N & 0.05 & 0.02 & 0.07 & 0.07 & 0.08 & 0.07 \\
\midrule
\multirow{2}{*}{v1.4} & Y & 0.98 & 0.67 & 0.96 & 0.98 & 0.98 & 0.98 \\
& N & 0.04 & 0.02 & 0.06 & 0.07 & 0.08 & 0.07\\
\bottomrule
\end{tabular}
}
\end{table}

\subsection{Transfer beyond DreamBooth}
\label{app:personalization-backends}
To test transfer beyond the main DreamBooth protocol, we evaluate LoRA-based fine-tuning~\cite{hu2021lora} and Textual Inversion~\cite{textual_inversion} on VGGFace2. Table~\ref{tab:backend_transfer} reports WM=Y/WM=N bit accuracies of $0.99/0.06$ for LoRA and $0.96/0.06$ for Textual Inversion; LoRA yields FID $185.90$. Both backends preserve a clear verification margin, suggesting that the semantic trace can transfer across personalization methods rather than depending solely on DreamBooth fine-tuning.

\begin{table}[t]
\small
\centering
\caption{Watermark transfer under different personalization backends on VGGFace2. DreamBooth is the main-paper default.}
\label{tab:backend_transfer}
\fitwidth{
\begin{tabular}{lccc}
\toprule
\textbf{Mimicry backend} & \textbf{WM=Y bit acc.} & \textbf{WM=N bit acc.} & \textbf{FID} $\downarrow$ \\
\midrule
DreamBooth~\cite{dreambooth} & 0.98 & 0.03 & 180.98 \\
LoRA~\cite{hu2021lora} & 0.99 & 0.06 & 185.90 \\
Textual Inversion~\cite{textual_inversion} & 0.96 & 0.06 & -- \\
\bottomrule
\end{tabular}}
\end{table}

\subsection{Multi-round adaptive generation and cherry-picking}
\label{app:multi-round-adaptive}
A stronger attacker may repeatedly generate candidates and publish only samples that minimize the watermark reader score while staying visually close to the target. We evaluate a multi-round generate-and-filter attacker: in each round, the attacker generates four candidates, retains the one with the lowest reader score under an identity-similarity budget of $0.92$ (with the current image also retained as a candidate), and repeats this process for three rounds. Table~\ref{tab:generate_filter} shows that FeatMark remains separable after this attack: WM=Y bit accuracy drops from $0.98$ to $0.72$, while WM=N remains $0.03$. The fidelity cost is also measurable, with FID increasing from $171.81$ to $192.13$, indicating that suppressing the semantic trace is not a free post-processing operation.

\begin{table}[t]
\small
\centering
\caption{Multi-round generate-and-filter attack on VGGFace2. The attacker performs three rounds, samples four candidates per round, and keeps the lowest-reader-score candidate under an identity-similarity budget of 0.92.}
\label{tab:generate_filter}
\fitwidth{
\begin{tabular}{lccc}
\toprule
\textbf{Setting} & \textbf{WM=Y bit acc.} & \textbf{WM=N bit acc.} & \textbf{FID} $\downarrow$ \\
\midrule
Default FeatMark & 0.98 & 0.03 & 171.81 \\
Generate-and-filter attacker & 0.72 & 0.03 & 192.13 \\
\bottomrule
\end{tabular}}
\end{table}

\subsection{Incremental real-world publishing}
\label{app:incremental-publishing}
In real deployments, a user may not have all future images available when first configuring a watermark. We therefore evaluate an incremental setting: FeatMark selects the message and feature program from an initial seed set of only two images and then applies the same program to later released images without re-selecting the feature. Table~\ref{tab:incremental_release} shows that this realistic variant remains close to the all-images-known setting, with WM=Y/WM=N bit accuracy of $0.98/0.04$ compared with $0.98/0.03$ by default. This suggests that feature selection does not require seeing every future release, provided that the selected semantic cue remains editable and identity/style-consistent for later images.

\begin{table}[t]
\small
\centering
\caption{Incremental publishing on VGGFace2. The seed-only setting selects the watermark program from two initial images and applies it to later releases.}
\label{tab:incremental_release}
\fitwidth{
\begin{tabular}{lcc}
\toprule
\textbf{Setting} & \textbf{WM=Y bit acc.} & \textbf{WM=N bit acc.} \\
\midrule
All images available at setup & 0.98 & 0.03 \\
Two-image seed set, later releases added & 0.98 & 0.04 \\
\bottomrule
\end{tabular}}
\end{table}

\subsection{Trigger-inversion, feature leakage, and forward security}
\label{app:trigger-inversion}
FeatMark should not be interpreted as a classic backdoor with a single universal trigger. Backdoor-recovery tools such as Neural Cleanse~\cite{neural_cleanse} assume that a compact, class-consistent trigger can be reverse-engineered by optimizing for a target behavior. In contrast, FeatMark's evidence is localized, identity/style-conditioned, and realized differently across samples. An attacker may infer a coarse candidate concept from the released image set, but does not observe the full defender-side program: the selected feature among semantically overlapping alternatives, the spatial grounding, the edit direction and strength, the private message-to-reader mapping, or the cloaking objective that transfers the cue through mimicry training. Consequently, concept neutralization can weaken some visible evidence but does not reliably erase the transferred semantic bias without rewriting identity/style details.

We also explicitly acknowledge the forward-security limitation. If an attacker learns the exact feature program and is willing to apply visible edits or discard many otherwise good generations, future attribution can be weakened. FeatMark is designed to raise the cost of such evasion and to expose a fidelity/identity tradeoff, not to provide perpetual cryptographic security after full feature leakage. This is why our evaluation includes targeted concept neutralization, diffusion regeneration, multi-round candidate filtering, and uncontrolled clean-image leakage.

\section{Ablation Study}
\label{sec:exp:ablation}
In this section, we evaluate image cloaking perturbation strength and an uncontrolled setting where the mimicry adversary obtains non-watermarked images. Generalization across diffusion model versions is reported in Appendix~\ref{sec:diffusion-generalization}.

\begin{table*}[t]
\small
\centering
\caption{Bit accuracy and perceptual quality of FeatMark in an uncontrolled setting with additional clean images used in mimicry attacks on VGGFace2. $\mathcal{N}_{\text{clean}}$ of 0 is the default setting (*) where the attacker can only access the cloaked images to perform mimicry attacks. The number of cloaked images provided to the attacker is 4 for all the results.}
\label{table:results_uncontrolled}
\fitwidth{
\begin{tabular}{ccccccccccccc}
\toprule
\multirow{2}{*}{$\mathcal{N}_{\text{clean}}$} 
& \multirow{2}{*}{\textbf{WM?}} 
& \multicolumn{6}{c}{\textbf{Bit Accuracy}} 
& \multirow{2}{*}{\textbf{FID} $\downarrow$}  
& \multirow{2}{*}{\textbf{Precision} $\uparrow$} 
& \multirow{2}{*}{\textbf{Recall} $\uparrow$} 
& \multirow{2}{*}{\textbf{Coverage} $\uparrow$} 
& \multirow{2}{*}{\textbf{Density} $\uparrow$} \\
\cmidrule(r){3-8}
& & None & Crop & Brigh. & Cont. & JPEG & Comb. & & & & & \\
\midrule
\multirow{2}{*}{0*} & Y & 0.98 & 0.75 & 0.96 & 0.96 & 0.99 & 0.98 & \multirow{2}{*}{180.98} & \multirow{2}{*}{0.47} & \multirow{2}{*}{0.19} & \multirow{2}{*}{0.83} & \multirow{2}{*}{0.62} \\
& N & 0.03 & 0.04 & 0.04 & 0.04 & 0.04 & 0.05 & & & & & \\
\midrule

\multirow{2}{*}{1} & Y & 0.97 & 0.73 & 0.96 & 0.96 & 0.97 & 0.96 & \multirow{2}{*}{180.12} & \multirow{2}{*}{0.53} & \multirow{2}{*}{0.24} & \multirow{2}{*}{1.03} & \multirow{2}{*}{0.75} \\
& N & 0.18 & 0.12 & 0.19 & 0.21 & 0.22 & 0.21 & & & & & \\
\midrule

\multirow{2}{*}{2} & Y & 0.98 & 0.68 & 0.95 & 0.96 & 0.98 & 0.97 & \multirow{2}{*}{177.95} & \multirow{2}{*}{0.53} & \multirow{2}{*}{0.32} & \multirow{2}{*}{1.06} & \multirow{2}{*}{0.79} \\
& N & 0.38 & 0.18 & 0.38 & 0.42 & 0.44 & 0.41 & & & & & \\
\midrule

\multirow{2}{*}{3} & Y & 0.99 & 0.69 & 0.97 & 0.98 & 0.99 & 0.97 & \multirow{2}{*}{176.34} & \multirow{2}{*}{0.57} & \multirow{2}{*}{0.30} & \multirow{2}{*}{1.15} & \multirow{2}{*}{0.81} \\
& N & 0.52 & 0.28 & 0.52 & 0.53 & 0.57 & 0.54 & & & & & \\
\midrule

\multirow{2}{*}{4} & Y & 0.98 & 0.64 & 0.95 & 0.96 & 0.97 & 0.95 & \multirow{2}{*}{174.89} & \multirow{2}{*}{0.59} & \multirow{2}{*}{0.38} & \multirow{2}{*}{1.15} & \multirow{2}{*}{0.83} \\
& N & 0.58 & 0.29 & 0.56 & 0.58 & 0.63 & 0.61 & & & & & \\
\midrule

\multirow{2}{*}{6} & Y & 0.98 & 0.63 & 0.94 & 0.96 & 0.97 & 0.95 & \multirow{2}{*}{173.50} & \multirow{2}{*}{0.64} & \multirow{2}{*}{0.31} & \multirow{2}{*}{1.41} & \multirow{2}{*}{0.87} \\
& N & 0.73 & 0.36 & 0.71 & 0.73 & 0.76 & 0.73 & & & & & \\
\midrule

\multirow{2}{*}{8} & Y & 0.99 & 0.68 & 0.97 & 0.98 & 0.99 & 0.99 & \multirow{2}{*}{172.10} & \multirow{2}{*}{0.59} & \multirow{2}{*}{0.39} & \multirow{2}{*}{1.26} & \multirow{2}{*}{0.87} \\
& N & 0.79 & 0.43 & 0.75 & 0.77 & 0.82 & 0.81 & & & & & \\
\bottomrule
\end{tabular}
}
\end{table*}

\begin{table*}[t]
\small
\centering
\caption{Bit accuracy and perceptual quality of FeatMark with different perturbation budget $\eta$, where $\eta=0.10$ is the default setting (*).}
\label{table:results_budget}
\fitwidth{
\begin{tabular}{ccccccccccccc}
\toprule
\multirow{2}{*}{$\eta$} 
& \multirow{2}{*}{\textbf{WM?}} 
& \multicolumn{6}{c}{\textbf{Bit Accuracy}} 
& \multirow{2}{*}{FID $\downarrow$}  
& \multirow{2}{*}{Precision $\uparrow$} 
& \multirow{2}{*}{Recall $\uparrow$} 
& \multirow{2}{*}{Coverage $\uparrow$} 
& \multirow{2}{*}{Density $\uparrow$} \\
\cmidrule(lr){3-8}
& & None & Crop & Brigh. & Cont. & JPEG & Comb. & & & & & \\
\midrule
clean & N & 0.03 & 0.04 & 0.04 & 0.04 & 0.04 & 0.05 & 171.81 & 0.65 & 0.28 & 1.33 & 0.83 \\
0.01 & Y & 0.25 & 0.15 & 0.27 & 0.26 & 0.25 & 0.23 & 173.92 & 0.60 & 0.32 & 1.19 & 0.80  \\
0.05 & Y & 0.85 & 0.59 & 0.83 & 0.88 & 0.81 & 0.80  & 178.46 & 0.43 & 0.27 & 0.75 & 0.61 \\
0.10* & Y & 0.98 & 0.75 & 0.96 & 0.96 & 0.99 & 0.98  & 180.98 & 0.47 & 0.19 & 0.83 & 0.62 \\
0.15 & Y & 0.99 & 0.81 & 0.99 & 0.99 & 0.99 & 0.98 & 183.57 & 0.43 & 0.18 & 0.80 & 0.59 \\
0.20 & Y & 0.99 & 0.84 & 0.99 & 0.99 & 0.99 & 0.99 & 185.21 & 0.43 & 0.18 & 0.73 & 0.56\\
\hline
\end{tabular}
}
\end{table*}

\noindent \textbf{Robustness under uncontrolled settings.~}
\label{sec:exp:uncontrolled}
Our prior settings are rigorously controlled to demonstrate the performance of FeatMark. We now consider a setting where the attacker not only has access to the cloaked images, but also additional clean images from the target prior to watermarking efforts.
Table~\ref{table:results_uncontrolled} is the result for our metrics of bit accuracy and perceptual quality with number of clean target samples ($\mathcal{N}_{\text{clean}}$) varied from 0 (default) to 8. 
When increasing $\mathcal{N}_{\text{clean}}$, the bit accuracy gap is reduced, with perceptual quality remaining higher, but FeatMark continue to effectively mitigate the mimic adversary. 
When, $\mathcal{N}_{\text{clean}}$=$2$ where $\mathcal{N}_{\text{clean}}$ is half of the number of cloaked images, FeatMark achieves bit accuracy with no transformation of 0.98-0.38. 
In the extreme case of $\mathcal{N}_{\text{clean}}$=$6$ and $\mathcal{N}_{\text{clean}}$=$8$ where $\mathcal{N}_{\text{clean}}$ is 1.5 or 2 times more populous than cloaked images, the accuracy gaps are too small to distinguish the watermark.

\noindent \textbf{Watermarking with different perturbation budgets.~}
We present in Table~\ref{table:results_budget} an ablation study when the perturbation budget, $\eta$ in Equation~\ref{eq:loss_cloak} is varied from 0.01 to 0.20 with 0.10 as the default setting in our prior experiments. 
Larger $\eta$ produces larger perturbations in the cloaked images. With each $\eta$ we record the corresponding bit accuracy and perceptual quality. 

From the results, a reduction in perturbation budget narrows the bit accuracy gap between watermarked and non-watermarked fake images, but improves the perceptual similarity between these images, making watermarks more difficult to be noticed. 
For example, when $\eta$ is 0.05, FeatMark preserves the bit accuracy with no transformation to 0.85-0.03, which is still large enough to distinguish watermarked and non-watermarked images, and achieve smaller FID of 178.46. 
In an extreme case where $\eta$ is 0.01, the perturbation has nearly no effect, so that bit accuracy gap drops to 0.25-0.03, which is not common and not recommended.

\section{Detection of Future Attacks}
\label{sec:discussion}

In this sub-section, we discuss potential threats in the future and argue how FeatMark remains effective against them.

\noindent \textbf{Stronger mimicry adversaries.~}
FeatMark's success relies on the mimicry capabilities of the adversary, a weak adversary generating poor fake representations of the target identity is unlikely to faithfully recreate inconspicuous feature watermarks. However, this setting is not a concern, as poorly crafted fakes are indicative of imitation.
On the other hand, stronger mimics will have a greater capacity to learn and replicate characteristics from the target's source data, in which FeatMark's visible features will also be present and transferred. 

\noindent \textbf{Visible watermarks raise the cost of attackers.~}
We have recently observed, claims of impossibly strong LLM watermarks~\cite{watermark_impossibility,can_ai_generated_text_be_reliably_detected} that are hard to remove provided adversaries are computationally bounded. 
Such a development may also occur in the text-to-image domain. 
Recent pixel based watermarks~\cite{stablesignature,diffusionshield,ft-shield,gaussian_shading,zodiac,promark,genwatermark} have been bypassed by simple perturbations~\cite{unmarker,wevade}. As we have shown, these simple perturbations cannot evade our watermark detection due to the clear presence of feature-level watermark in the image. Unlike visible text watermarks~\cite{advdm, visible_watermark}, FeatMark is naturally integrated into the image, alleviating the problem of image quality degradation. Already, FeatMark shows that watermarked features vary their location and specific forms across multiple fake images. 
With no clear position or form, the cost of removal is expected to increase beyond the cost of existing attacks against pixel-level watermarks.

\end{document}